\documentclass[journal=jacsat,manuscript=article]{achemso}
\usepackage{chemformula} 
\usepackage[T1]{fontenc} 
\usepackage{subcaption}
\usepackage{hyperref}
\usepackage{xcolor}

\newcommand{\mycomment}[1]{%
  {\color{black}#1}%
}

\title{Fewest switches surface hopping \mycomment{with decoherence} in the Marcus inverted regime: correct rates but wrong thermal populations}

\author{ Manas Nagda, Priyam Kumar De, and Amber Jain}

\affiliation{Department of Chemistry, Indian Institute of Technology Bombay, Powai, Mumbai-400076, India}
\email{amberj@chem.iitb.ac.in}

\date{\today}

\begin{document}


\begin{abstract}
Fewest switches surface hopping (FSSH) is a well benchmarked dynamical method for simulating nonadiabatic systems. In particular, the literature shows that for the spin-Boson model Hamiltonian, FSSH with appropriate corrections usually captures the detailed balance well and obtains rate constants within a factor of 2 compared to numerically exact results. In this study, we show that in the deep inverted Marcus regime, \mycomment{the augmented-FSSH (AFSSH, one version that includes decoherence)}  yields reasonably accurate rate constants but incorrect thermal populations over a broad range of parameters. We present an analytical derivation to understand the AFSSH behavior, and therefore, show that AFSSH obtains correct rate constants owing to the resonance of the time derivative coupling with the exothermicity, \mycomment{but} obtains an incorrect thermal population owing to the self-consistency issue. The presented derivation provides an analytical expression for the quantum correction factor for AFSSH simulations in the Marcus inverted regime.


\end{abstract}

Fewest switches surface hopping (FSSH)\cite{tully1990molecular} is a popular method for non-adiabatic simulations for a wide range of applications.\cite{chapman1992classical,jasper2011non,barbatti2011nonadiabatic,malhado2014non,subotnik2016understanding,gonzalez2020quantum}  With appropriate treatment of over-coherence and frustrated hops, FSSH performs well for several model Hamiltonians\cite{muller1997surface,dou2015surface,plasser2019strong} as well as molecular benchmarks.\cite{barbatti2007fly,aarti2021,de2024exciton} \mycomment{In this work, we use the augmented-FSSH (AFSSH)\cite{landry2012recover}, which is one of the variants of decoherence.} There are some cases known when rate constants can be much worse from \mycomment{AFSSH},\cite{schwartz1996quantum,prezhdo1997evaluation,wong2002solvent,lawrence2024recovering} however, the thermal populations from \mycomment{AFSSH} remain accurate with very few exceptions.\cite{parandekar2005mixed,schmidt2008mixed,jain2015surface,sherman2015thermal,sifain2016communication,carof2017detailed}

\mycomment{The main aim of this work is to investigate the applicability of AFSSH in the deep inverted Marcus regime where the exact rate constants are orders of magnitude larger than the Marcus rate constants.\cite{siders1981quantum,liang1990temperature,menzeleev2011direct,lawrence2018analytic,heller2020instanton} Given that AFSSH captures the Marcus rate constants well,\cite{landry2012recover,jain2015surface,jain2015does,richardson2024nonadiabatic} we investigate if AFSSH is applicable for weak reorganization energies where Marcus theory fails. 

Deep inverted Marcus regime features nearly parallel energy surfaces where the diabatic crossing point is thermally inaccessible and several problems belong to this category.\cite{liang1990temperature,xi2025revisiting} Photosynthetic systems, such as FMO,\cite{ishizaki2009theoretical} have parameters in the inverted regime and some discrepancies were observed between AFSSH and numerically exact results for the FMO problem.\cite{sindhu2022coherence,sindhu2024electronic} Another important class of problems is where a vibrational mode is explicitly quantized in AFSSH simulations, leading to nearly parallel vibronic surfaces.\cite{hammes1994proton,kab2006fewest,jain2018vibrational,aarti2021} For such problems where non-adiabatic hops occur close to the minima, rather than at the diabatic crossing points, its worth investigating if AFSSH can capture the correct dynamics.
}

\mycomment{Importantly, we show in this work} that the thermal populations obtained from AFSSH can be significantly incorrect in the deep-inverted Marcus regime (i.e., with weak system-bath coupling). Interestingly, AFSSH obtains rate constants within a factor of two compared to the numerically exact results in this regime \mycomment{(while Marcus rate constants are orders of magnitude smaller)}.



We provide a detailed analytical formula that combines decoherence, frustrated hops, and inconsistency issues\cite{granucci2007critical,wang2014global,sifain2015mixed} of the AFSSH method to quantify its unusual behavior for the systems and parameters investigated in this study. We show that AFSSH obtains \mycomment{correct trend of the} rate constants \mycomment{for the spin-Boson Hamiltonian} for the correct reasons: the rate is dictated by the vibrational mode in resonance with the exothermicity. This resonance is correctly captured by the AFSSH method in the adiabatic basis, \mycomment{particularly for a harmonic bath}. The lack of detailed balance in AFSSH arises from the inconsistency issue, i.e., the populations computed using quantum coefficients and the fraction of trajectories on a given surface do not match. The inconsistency is a result of the slow decoherence timescale because the potential energy surfaces are nearly parallel, leading to an inaccurate quantum correction factor.


The Hamiltonian chosen for this work is the well studied spin-Boson model, with the Hamiltonian given by
\begin{align}\label{eq_H}
    H = \sum_i \frac{p_i^2}{2m} +\begin{bmatrix}
   \sum_{i} \frac{1}{2} m \omega_{i}^{2} x_{i}^{2} & V_{c} \\
    V_{c} & \Delta G + \sum_{i} \frac{1}{2} m \omega_{i}^{2}(x_{i} + \frac{g_{i}}{m\omega_{i}^{2}})^{2}
\end{bmatrix},
\end{align}
where $\Delta G$ is the exothermicity, and $V_c$ is the diabatic coupling. The frequency, coordinate, and momentum of the $i$\textsuperscript{th} bath mode are given by $\omega_i$, $x_i$ and $p_i$, respectively. The system-bath coupling is described by the Drude spectral density: 
\begin{equation}
    J(\omega)\equiv \frac{\pi}{2} \sum_{i} \frac{g_i^2}{m\omega_i}\delta(\omega-\omega_i)= 2\lambda\frac{\eta \omega}{\eta^2+ \omega^2},
    \label{eq_spec}
\end{equation}
where $g_i$ is the coupling strength of the system with the $i$\textsuperscript{th} bath mode. $\lambda$ is the reorganization energy and $\eta$ is the Drude characteristic frequency. 

\mycomment{The specific parameters chosen here are motivated from the FMO system.\cite{ishizaki2009theoretical} These are known to be in weak reorganization energy limit. Particularly, the parameters are close to the sites 2-3 of the FMO system which is in the inverted Marcus regime. We have chosen a temperature of 300 K (instead of the more common 77 K for the FMO problem) since we are treating all vibrations classically and reorganization energy of 10 cm$^{-1}$, leading to a barrier height of $2890$ cm$^{-1}$ because we are interested in the deep inverted regime (the usual reorganization energy is 70 cm$^{-1}$, which gives barrier height of $280$ cm$^{-1}$).\cite{aarti2021}}
All chosen parameters are in the deep inverted Marcus regime, and the standard value along-with the range of parameter values investigated is provided in Table \ref{table_param}. 

\begin{table}[ht]
    \centering
    \begin{tabular}{c|c|c}
        Parameter & Value & Range \\
        \hline
        Mass $(m)$ & 1837 a.u. & - \\
        Exothermicity ($\Delta G$) & -350 cm$^{-1}$ & -400 cm$^{-1}$ to -250 cm$^{-1}$\\
        Reorganization energy $(\lambda)$ & 10 cm$^{-1}$ & 5 cm$^{-1}$ to 20 cm$^{-1}$\\
        Drude Characteristic frequency $(\eta)$ & 50 cm$^{-1}$ & -\\
        Diabatic coupling ($V_c$) &30 cm$^{-1}$ & -\\
        Temperature (T) & 300 K & 50 K to 600 K\\
    \end{tabular}
    \caption{Values of the parameters defining the model Hamiltonian and the range of the parameters used in the study.}
    \label{table_param}
\end{table}


The FSSH method has been extensively reviewed in the literature.\cite{barbattiwires,subotnik2016understanding,jain2022pedagogical} Briefly, FSSH is a mixed quantum-classical \mycomment{(MQC)} method. The electronic wave function is expanded in the adiabatic basis
\begin{equation}
    |\Psi(\vec{x},t)\rangle = \sum_{i}c_i^{ad}(t)| \psi_{i}(\vec{x})\rangle,
\end{equation}
where $c_i^{ad}(t)$ is the time-dependent complex-valued coefficient and $|\psi_{i}(\vec{x})\rangle$ is the $i$\textsuperscript{th} adiabatic state depending on the classical coordinates $\vec{x}$. 

Using the electronic wavefunction in the time-dependent Schrödinger equation (TDSE), we obtain 
\begin{equation}
    i\hbar \dot{c}_j^{ad}(t) =E_j(\vec{x})c_j^{ad}(t) -\sum_k i\hbar {\vec{v}}.\vec{d}_{jk}c_k^{ad}(t),
    \label{eq_cdot}
\end{equation}
where $E_j$ is the adiabatic eigen energy, $\vec{v}$ is the velocity vector, and the time derivative coupling between adiabatic states $j$ and $k$ is given by ${\vec{v}}.\vec{d}_{jk} = \langle \psi_j\vert \frac{\partial\psi_k}{\partial t}\rangle$. 

The surface hopping algorithm evolves the classical  degrees of freedom ($\vec{x}$) using Newtonian mechanics, typically using an adiabatic PES. The hopping probability from adiabatic surface $j$ to $k$ is calculated from the complex coefficients as 
\begin{equation}\label{hop_prob}
    g_{jk} = -\frac{2 Re(a^*_{jk}{\vec{v}}.\vec{d}_{jk})dt}{a_{kk}}
\end{equation}
where $a_{jk} = (c_j^{ad})^*c_k^{ad}$ and $dt$ denotes the classical time step. 

The FSSH method tends to accumulate over-coherence errors, and decoherence corrections are added to the FSSH when the trajectory is far from the coupling region, where wave packets on different surfaces have a small overlap. Over the past few decades, several theoretical groups have proposed various decoherence schemes \cite{hammes1994proton,bittner1995quantum,schwartz1996quantum,prezhdo1997evaluation,jasper2002fewest,granucci2010,subotnik2011new,jain2016efficient}. In this study, we use the AFSSH method developed by the Subotnik group.\cite{landry2012recover} Other discrepancies arising from velocity reversal and calculating populations can be found in detailed reviews.\cite{subotnik2016understanding,jain2022pedagogical}  Our codes for AFSSH dynamics are available at https://github.com/mnagda/AFSSH-dripping.git.


We compare the AFSSH results with those obtained using the numerically exact hierarchical equations of motion (HEOM) method. The details of the HEOM method can be found in Refs. \citenum{tanimura1989time, ishizaki2005quantum,tanimura2020numerically,chen2009optical} and our codes are freely available at https://github.com/amber-jain-group-iitb/HEOM-generalized.git. 
 For all the parameters chosen for this work, the population decay in the diabatic basis fits well to an exponential decay:
\begin{align}
    \dot{P}_\mycomment{2}=-k_f P_\mycomment{2} + k_b P_\mycomment{1},
    \label{eq_rate_model}
\end{align}
where $P_1$ ($P_2$) is the population of diabat 1 (diabat 2), and diabat 1 is the \mycomment{lower}-energy surface. Fitting the population as a function of time provides the rate constants and long-time populations for both the AFSSH and HEOM methods. The plots below show the total rate constant $k=k_f+k_b$ and the long-time population $k_b/k$ (of the high-energy state). The FSSH results without decoherence do not follow an exponential decay, and the population plot for FSSH is presented later in this paper. The AFSSH and HEOM simulation details are provided in the SI.


\mycomment{Figure \ref{pes_dyn} (a) shows the adiabatic PES for the standard values of the parameters given in the column 2 of Table \ref{table_param}. It also shows the variation of nonadiabatic coupling vector along the reaction coordinate.}
The populations of diabat \mycomment{2} calculated from the AFSSH simulation, Marcus theory, and HEOM simulation for the standard parameters are shown in Fig. \ref{pes_dyn} (b). \mycomment{The standard Marcus  rate constant is given by}\cite{Nitzan}
\begin{align}
    \mycomment{k_{Marcus} = \frac{\vert V_c \vert ^2}{\hbar} \sqrt{\frac{\pi}{\lambda k_B T}}\  \textrm{exp} \left[ \frac{-(\Delta G \pm\lambda)^2}{4 \lambda k_B T} \right]}
    \label{eq_k_Marc}
\end{align}
Clearly, the AFSSH rates are closer to the HEOM rates, whereas the long-time population from AFSSH is far from the Boltzmann population of $\sim$  \mycomment{0.16, shown as the dashed line in this figure}. The large difference between the AFSSH and Marcus rates further shows that the transitions in the AFSSH are governed by trajectories directly dripping down from surface \mycomment{2} to \mycomment{1} without accessing the diabatic crossing point. \mycomment{To show this point clearly, red dots in Fig. \ref{pes_dyn} (a) show the position and energy where a hop occurred. The vast majority of hops are around the minima ($8471$ hops out of $10000$ within $4k_BT$ energy of minima) and not near the diabatic crossing point.}

\mycomment{It is intriguing that the AFSSH rate constants match better with HEOM results than with the Marcus theory, given multiple works in literature have shown that the AFSSH rate constants match reasonably well with Marcus theory.\cite{landry2012recover,jain2015surface,jain2015does,richardson2024nonadiabatic} To understand this mismatch, consider the Fermi's golden rule (FGR) rate constant in the diabatic basis:
\begin{align}
    \mycomment{k_{FGR}^{d} = \frac{\vert V_c\vert ^2}{\hbar}\int_{-\infty}^{\infty} dt \ e^{-i \Delta G t/\hbar} \textrm{exp} \left\{ -\frac{1}{\hbar}\int_0^\infty d\omega \frac{J(\omega)}{\pi\omega^2} \left(\textrm{coth}\left( \frac{\beta \hbar \omega}{2}\right) (1-\cos(\omega t) ) - i \sin\omega t \right) \right\}.}
    \label{eq_k_FGR}
\end{align}

Equation \ref{eq_k_FGR} simplifies to the Marcus rate constant Eq. \ref{eq_k_Marc} in the high temperature limit [i.e. $\tanh(0.5\beta\hbar\omega)\approx \beta \hbar\omega/2$] and small time limit [i.e $\sin(\omega t)\approx \omega t$ and $(1-\cos(\omega t))\approx 0.5\omega^2t^2$]. For the parameters investigated here, the small time limit is not valid. If only the high temperature approximation is made, we obtain
\begin{align}
    \mycomment{k_{FGR}^{d,T\uparrow} \approx \frac{\vert V_c\vert ^2}{\hbar}\int_{-\infty}^{\infty} dt \ e^{-i \Delta G t/\hbar} \textrm{exp} \left\{ -\frac{1}{\hbar}\int_0^\infty d\omega \frac{J(\omega)}{\pi\omega^2} \left(\left( \frac{2}{\beta \hbar \omega}\right) (1-\cos(\omega t) ) - i \sin\omega t \right) \right\}.}
    \label{eq_k_FGR_highT}
\end{align}

\begin{figure}[H]
    \begin{subfigure}{0.90\linewidth}
        \centering
        \includegraphics[width=\linewidth]{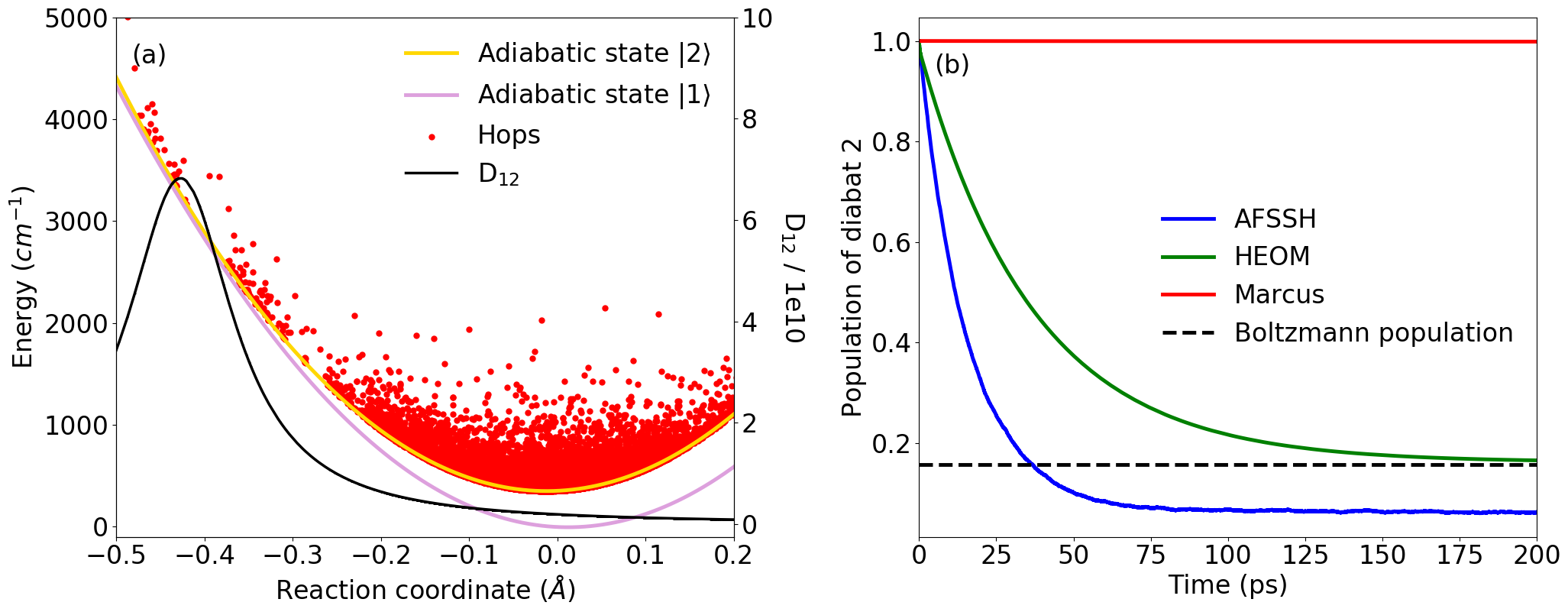}
        \label{fig_1a}
    \end{subfigure}
    \begin{subfigure}{0.425\linewidth}
        \centering
        \includegraphics[width=\linewidth]{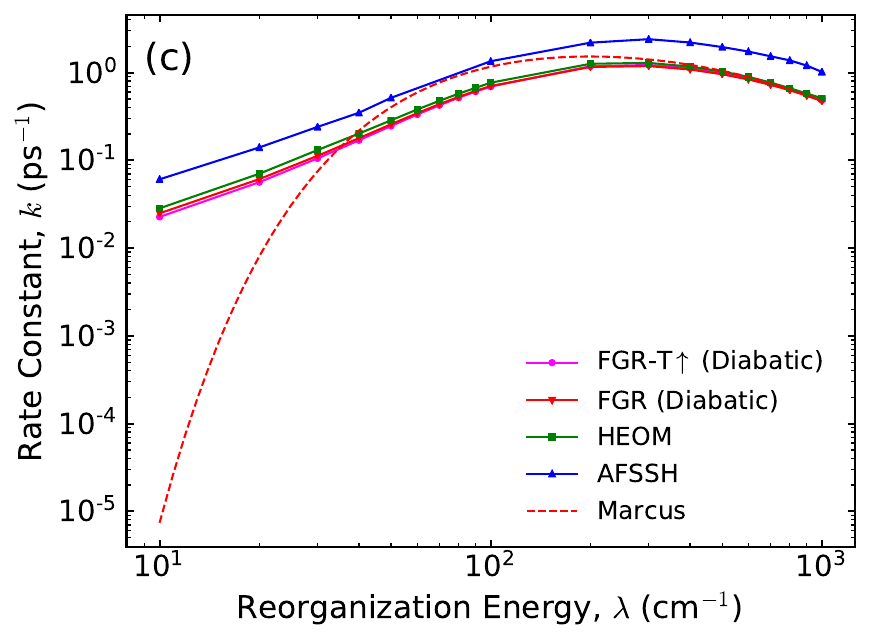}
        \label{rate_comp_all}
    \end{subfigure}
    \quad
    \begin{subfigure}{0.425\linewidth}
        \centering
        \includegraphics[width=\linewidth]{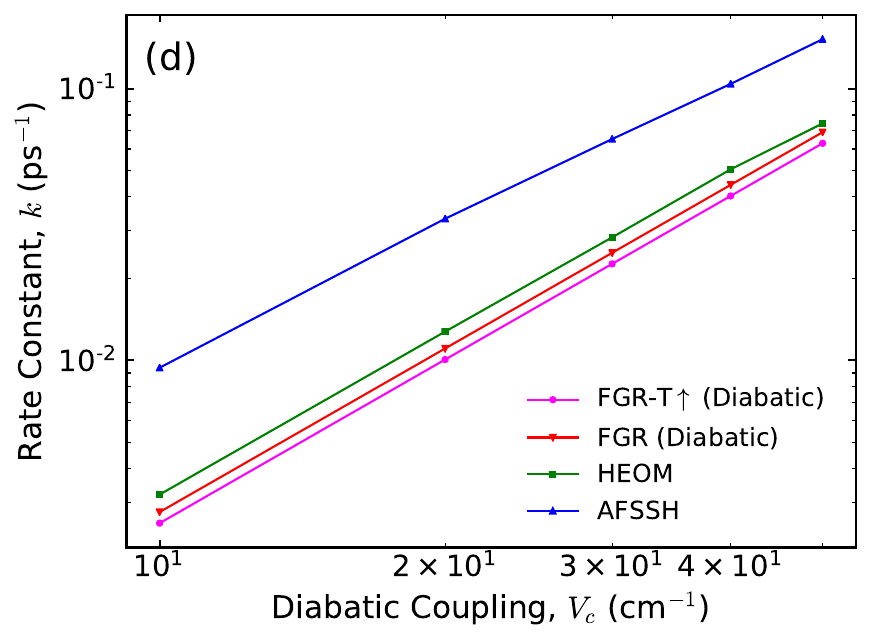}
        \label{rate_Vc_AFSSH_HEOM}
    \end{subfigure}
    \caption{(a) Electronic adiabatic PES and nonadiabatic coupling vector as a function of classical coordinate. \mycomment{The red dots correspond to the position of hops between two states along the reaction coordinate.} (b) Long time diabatic population dynamics for parameters of column 2 of Table \ref{table_param} obtained from HEOM, AFSSH and Marcus rate theory. \mycomment{Boltzmann population is shown as a dashed black line.} \mycomment{(c) Comparison of rate constants obtained from high temperature FGR (Eq. \ref{eq_k_FGR_highT}), FGR formula (Eq. \ref{eq_k_FGR}), HEOM, AFSSH and Marcus formula (Eq. \ref{eq_k_Marc}) for the variation of reorganization energy. (d) Comparison of rate constants obtained from high temperature FGR theory, FGR theory, HEOM and AFSSH for variation in diabatic coupling constant.}}
    \label{pes_dyn}
\end{figure}

%



Rate constants computed from AFSSH, HEOM, FGR (Eq. \ref{eq_k_FGR}) and FGR with only the high temperature limit (Eq. \ref{eq_k_FGR_highT}) are compared with varying $\lambda$ in Fig. \ref{pes_dyn} (c) and with varying $V_c$ in Fig. \ref{pes_dyn} (d). All other parameters are the standard parameters. The Marcus rate constant (Eq. \ref{eq_k_Marc}) is put for comparison in Fig. \ref{pes_dyn} (c), which is 3 orders of magntiude smaller than FGR rate constant for $\lambda=10$ cm$^{-1}$. It can be seen that high temperature approximation is good across entire range of $\lambda$, but small time limit breaks for small reorganization energies ($\lambda<50$ cm$^{-1}$). The integrands in Eq. \ref{eq_k_FGR} are explicitly shown in Fig. S3 of SI to show the breakdown of the small time approximation.

Instead of taking small time limit, if small $\lambda$ limit is taken, we show in SI that Eq. \ref{eq_k_FGR} can be simplified to
\begin{align}
k_{FGR}^{\mycomment{2}\rightarrow \mycomment{1}} &= \frac{2k_BTV_c^2}{\hbar |\Delta G|^3}J\left(\frac{|\Delta G|}{\hbar}\right) \frac{\beta|\Delta G|}{1-e^{-\beta|\Delta G|}},\label{eq_LT_limit1}
\\
k_{FGR}^{\mycomment{1}\rightarrow \mycomment{2}} &= \frac{2k_BTV_c^2}{\hbar |\Delta G|^3}J\left(\frac{|\Delta G|}{\hbar}\right) \frac{\beta|\Delta G|}{1-e^{-\beta|\Delta G|}}e^{-\beta|\Delta G|}.
\label{eq_LT_limit2}
\end{align}
Here, $\frac{\beta\Delta G}{1-e^{-\beta\Delta G}}$ is sometimes referred to as the harmonic quantum correction factor from a Landau-Teller perspective.\cite{bader1994quantum} 

Equations \ref{eq_LT_limit1} and \ref{eq_LT_limit2} can also be derived with FGR applied in adiabatic basis, treating the derivative coupling $\sum_j i\hbar \hat{p}_jd_{j}^{12}/m$ as the perturbation (with $\hat{p}_j$ being the momentum operator of the $j$\textsuperscript{th} bath mode). For completeness, we have provided a derivation in the SI. The important assumptions made are: (a) $d_j^{12}$ is assumed to be a constant, and (b) $\langle j_k^I | j_k^J\rangle\approx 1$ where $|j_k^I\rangle$ and $|j_k^J\rangle$ represent vibrational eigen functions of the electronic adiabatic states $I$ and $J$, respectively.

We will show in this paper that AFSSH rate constants are proportional to the FGR rate constants, and is the reason why AFSSH captures the correct trends in the rate constants.

Equations \ref{eq_LT_limit1} and \ref{eq_LT_limit2} are not fundamentally new results. These are closed form expressions for the semiclassical limit for the adiabatic rate constants derived by Bhattacharyya {\it et al.}\cite{bhattacharyya2013adiabatic} and recently by Rhee {\it et al.}\cite{jang2025fermi} Same results also appear from the modified Redfield theory.\cite{yang2002influence,hwang2015coherent,upadhyay2025electronic} From a different perspective, path integral approaches have also been used to calculate the rate constants for the Marcus inverted regime.\cite{egger1994quantum,menzeleev2011direct,lawrence2018analytic,heller2020instanton,richardson2024nonadiabatic}


The Marcus rate constant Eq. \ref{eq_k_Marc} and the rate constant of Eq. \ref{eq_LT_limit1} (or \ref{eq_LT_limit2}) present two different mechanisms of population transfer. Marcus rate constant is interpreted as population transfer through the diabatic crossing, with a transition state mechanism. In contrast, Eq. \ref{eq_LT_limit1} allows a direct population transfer close to the energy surface minima when the energy surfaces are nearly parallel (small $\lambda$ limit). The transfer is driven by the bath mode in resonance with the exothermicity, and the Franck-Condon overlap around the minima. 

To investigate the deep inverted regime, the parameters chosen in Table \ref{table_param} are where rate constant of  Eq. \ref{eq_LT_limit1} is greater (typically an order of magnitude larger) than the Marcus rate constant. To the best of our knowledge, surface hopping has not been benchmarked in the regime where Marcus and FGR results deviate significantly.}

\mycomment{Moving on to interpret the AFSSH results,} we
build a model with correct qualitative trends. We make the following main assumptions: (a) deep inverted regime: dynamics occur close to the minima and the crossing region is thermally inaccessible. Further, we assume the derivative coupling is weak and can be treated perturbatively. \mycomment{This assumption was also made in deriving the FGR rate constants (Eqs. \ref{eq_LT_limit1} and \ref{eq_LT_limit2}).}  (b) Nuclear velocities are classical and are always at thermal equilibrium. (c) A constant decoherence timescale ($\tau$) is assumed which is much smaller than the population decay timescale, but much larger than the vibrational energy relaxation timescale. 

Before deriving the AFSSH rate constants, we show the \mycomment{MQC} Fermi's golden rule (FGR) rate constant for population transfer between adiabatic surfaces \mycomment{2} and \mycomment{1}  (surface \mycomment{2} is higher in energy compared to surface \mycomment{1}) with the coupling $i\hbar\vec{v}.\vec{d}_{12}$ treated as a perturbation:
\begin{align}
    k_{FGR,\mycomment{MQC}}^{\mycomment{2}\rightarrow \mycomment{1}} = \sum_{j} \frac{V_{c}^{2}}{2\langle\delta V_{ad}^+\rangle^{4}}\frac{k_{B}T}{m} g_{j}^{2}\sqrt{\frac{\pi\hbar^{2}}{\lambda k_{B}T}} \exp\left(\frac{-(\Delta G + \lambda + \hbar\omega_{j})^{2}}{4\lambda k_{B}T}\right),
    \label{k_fgr_no_deco}
\end{align}
where $\langle\delta V_{ad}^+\rangle=\Delta G+\lambda$ and $k_B$ is the Boltzmann constant.
Similarly, the \mycomment{MQC} FGR rate constant for the $\mycomment{1}\rightarrow \mycomment{2}$ adiabatic transition is given by 
\begin{align}
    k_{FGR,\mycomment{MQC}}^{\mycomment{1}\rightarrow \mycomment{2}} = \sum_{j} \frac{V_{c}^{2}}{2\langle\delta V_{ad}^-\rangle^{4}}\frac{k_{B}T}{m}g_{j}^{2} \sqrt{\frac{\pi\hbar^{2}}{\lambda k_{B}T}} \exp\left(\frac{-(\Delta G - \lambda + \hbar\omega_{j})^{2}}{4\lambda k_{B}T}\right)
    \label{k_fgr_reverse}
\end{align}
with $\langle\delta V_{ad}^-\rangle=\Delta G-\lambda$. A detailed derivation is provided in the SI. \mycomment{Equations \ref{k_fgr_no_deco} and \ref{k_fgr_reverse} are same as Eq. 37 of Ref. \citenum{renger2021semiclassical}, which derived a semiclassical rate expression using modified Redfield theory. As a sanity check, the limit $\lambda\rightarrow 0$, using Eq. \ref{eq_spec}, Eqs. \ref{k_fgr_no_deco} and \ref{k_fgr_reverse} transform to}
\begin{align}
\lim_{\lambda\rightarrow0}k_{FGR,MQC}^{1\rightarrow2} =\lim_{\lambda\rightarrow0}k_{FGR,MQC}^{2\rightarrow1}= \frac{2k_BTV_c^2}{\hbar |\Delta G|^3}J\left(\frac{|\Delta G|}{\hbar}\right),
\label{eq_kfgr_0lambda}
\end{align}
\mycomment{which is the same as Eqs. \ref{eq_LT_limit1} and \ref{eq_LT_limit2} with the quantum correction factor missing. This should not be surprising: mixed quantum-classical approaches can miss detailed balance, and quantum correction factors are often required to account for detailed balance.\cite{bader1994quantum}}


The AFSSH rate constants deviate from the FGR rate constants owing to over-coherence and frustrated hops. 
We present a simple model in the SI of the AFSSH rate constants, assuming  a constant decoherence timescale $\tau$. The final rate expressions for the AFSSH rate constants are
\begin{align}
\label{k_rate_final}
k_f^{AFSSH}&=k_{FGR,\mycomment{MQC}}^{\mycomment{2}\rightarrow \mycomment{1}}  \left(\frac{k_{c_{\mycomment{2}}=1}^{\mycomment{1}\rightarrow \mycomment{2}} \tau }{1+k_{c_{\mycomment{2}}=1}^{\mycomment{1}\rightarrow \mycomment{2}} \tau }\right) e^{\beta |\Delta G|},\\
    \label{k_rate_final1}
    k_{b}^{AFSSH} &= k_{FGR,\mycomment{MQC}}^{\mycomment{1}\rightarrow \mycomment{2}}{\left(\frac{k^{\mycomment{2}\rightarrow \mycomment{1}}_{c_{\mycomment{1}}=1}\tau}{1+ k^{\mycomment{2}\rightarrow \mycomment{1}}_{c_{\mycomment{1}}=1}\tau}\right)}e^{-\beta |\Delta G|},
\end{align}
with
\begin{align}
k^{\mycomment{1}\rightarrow \mycomment{2}}_{c_{\mycomment{2}} = 1} &=  \frac{4g\left\langle|v|\right\rangle}{\langle|\delta V_{ad}^+|\rangle}e^{-\beta |\Delta G|},\\
k^{\mycomment{2}\rightarrow \mycomment{1}}_{c_{\mycomment{1}} = 1}& =\frac{4g\left\langle|v|\right\rangle}{\langle|\delta V_{ad}^-|\rangle}.
\end{align}
Here $g=\sqrt{2m\omega_0^{2}\lambda}$, $\langle|v|\rangle=\sqrt{\frac{k_B T}{2\pi m}}$, and $\omega_0$ is the frequency of the reaction coordinate.

Equations \ref{k_rate_final} and \ref{k_rate_final1} are the main results of this paper to understand \mycomment{AFSSH} rate constants in the deep inverted regime. $k_{FGR,\mycomment{MQC}}^{\mycomment{2}\rightarrow \mycomment{1}}$ and $k_{FGR,\mycomment{MQC}}^{\mycomment{1}\rightarrow \mycomment{2}}$ are calculated using Eqs. \ref{k_fgr_no_deco} and \ref{k_fgr_reverse}. In practice, Eq. \ref{eq_kfgr_0lambda} can be used as well for computational ease, and the results shown later remain roughly the same if Eq. \ref{eq_kfgr_0lambda} is used instead of Eqs. \ref{k_fgr_no_deco} and \ref{k_fgr_reverse}. Using Eq. \ref{eq_kfgr_0lambda} also provides an analytical expression for the quantum correction factor for AFSSH simulations. Equations \ref{k_rate_final} and \ref{k_rate_final1} can be rewritten as
\begin{align}
    k_{i\rightarrow j}^{AFSSH} = \left(\lim_{\lambda\rightarrow 0}k_{FGR,\mycomment{MQC}}^{i\rightarrow j}\right) Q(\Delta G_{i\rightarrow j}),
    \label{eq_kAFSSH_QCF}
\end{align}
where $\Delta G_{\mycomment{2}\rightarrow \mycomment{1}}=-\Delta G$ and $\Delta G_{\mycomment{1}\rightarrow \mycomment{2}}=\Delta G$, and the quantum correction factor given by
\begin{align}
    Q(\Delta G_{i\rightarrow j})=\frac{\langle k \rangle e^{-\beta \Delta G_{i \rightarrow j}/2} }{1+\langle k \rangle e^{-\beta \Delta G_{i \rightarrow j}/2}}.
\end{align}
Collecting terms from Eqs. \ref{k_rate_final} and \ref{k_rate_final1}, $\langle k \rangle= \frac{4\omega_0 \tau}{\sqrt{\pi}}\frac{\sqrt{\lambda k_B T}}{|\Delta G|}e^{-\beta |\Delta G|/2}$.

The decoherence timescale 
 $\tau$ is required to evaluate Eqs. \ref{k_rate_final} and \ref{k_rate_final1}, which can be estimated from any method of user's choice. In this work, we estimate $\tau$ by simulating 1000 AFSSH trajectories for 10 ps for each parameter and calculating the average collapse time. The values of $\tau$ calculated in this way are shown in Table S1 of SI for all the parameters.

\begin{figure}[ht]
    \centering
    \includegraphics[width=0.98\linewidth]{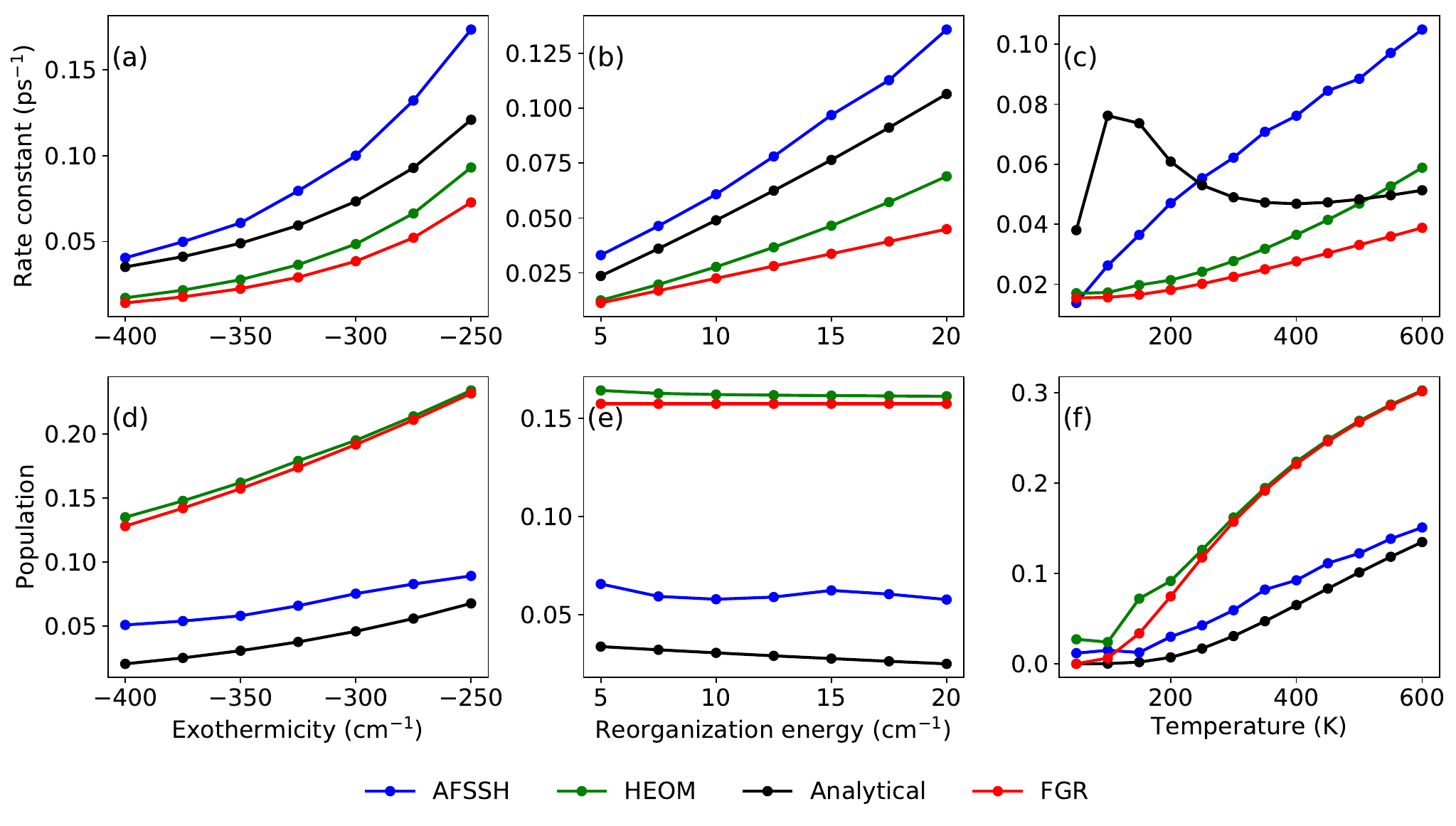}
    \caption{Comparison of the rate constants and long time populations obtained from AFSSH, HEOM, analytical rate theory (Eqs. \ref{k_rate_final} and \ref{k_rate_final1}) and FGR (Eqs. \ref{eq_LT_limit1} and \ref{eq_LT_limit2}) as a function of (a) exothermicity, (b) reorganization energy, and (c) temperature with parameters taken from Table \ref{table_param}. Lower panels shows comparison of the thermal population while upper panel show the rate constants.}
    \label{rate_population}
\end{figure}

Figure \ref{rate_population} shows the rate constants and long-time population computed using AFSSH, HEOM, \mycomment{FGR theory (Eqs. \ref{eq_LT_limit1} and \ref{eq_LT_limit2})} and our analytical rate expressions (Eqs. \ref{k_rate_final} and \ref{k_rate_final1}) for the range of parameters listed in Table \ref{table_param}. The rate constants from the AFSSH (Fig. \ref{rate_population} (a)-(c)) are roughly a factor of 2 larger than the HEOM results and our analytical rate constants are also within a factor of 2 of the AFSSH results. The AFSSH long time populations (Fig. \ref{rate_population} (d)-(f)) are significantly smaller than the correct HEOM results. Our analytical expressions correctly capture the qualitative trends of most of the AFSSH results. \mycomment{The FGR rates are close to the HEOM results for the most part. Some deviations at high temperatures and high $\lambda$ can be seen where the Marcus rate constants start becoming comparable to the FGR rate constants (of Eqs. \ref{eq_LT_limit1} and \ref{eq_LT_limit2}).}

The most striking aberration of our analytical formula is seen in Fig. \ref{rate_population} (c): rate constants as a function of temperature. While both the AFSSH and HEOM rate constants increase monotonically with temperature, the analytical results show an erroneous maximum at approximately 150 K. We believe that a more accurate description of decoherence needs to be included in our formula to correctly capture the temperature trend. However, we note that our formula goes to the correct limit at $T= 0$ K, and the predictions of the thermal populations remain reasonably accurate for all parameters.

Our analytical expressions provide an immediate explanation for why the AFSSH long-time populations are much smaller than the true Boltzmann answer. Consider the limit of large $\tau$, or specifically when $k^{2\rightarrow1}_{c_{1} = 1}\tau\gg 1$ and $k^{1\rightarrow2}_{c_{2} = 1}\tau\gg 1$ (which is true for most of the parameters in the inverted regime), using Eqs. \ref{k_rate_final} and \ref{k_rate_final1}
\begin{align}
    \frac{k_b^{AFSSH}}{k_f^{AFSSH}}\approx e^{-2\beta \Delta G}.
\end{align}
 This shows that for a slow decoherence time scale, the AFSSH obtains thermal equilibrium at half the temperature of the true answer.

Next, we discuss why the AFSSH rate constants are close to the HEOM rate constants rather than the Marcus rate constants. \mycomment{Equation \ref{eq_kAFSSH_QCF} shows that AFSSH results are proportional to the FGR rate constants, with an incorrect quantum correction factor. Therefore, AFSSH rate constants follow the correct trends with wrong thermal populations.

We expect the rate constant agreement to be particularly better for a harmonic bath.} The AFSSH captures the resonance of the vibrational mode with exothermicity via the non-adiabatic coupling term \mycomment{$\vec{v}(t).\vec{d}_{12}\approx\sum_k v_k d_k^{12} \cos(\omega_kt)$. This term acts as an external perturbation to the quantum amplitudes (see Eq. \ref{eq_cdot}), and in the weak $\lambda$ regime the resonance condition is $\omega_k=\Delta G/\hbar$. For a harmonic bath, the classical resonance condition of AFSSH is the same as the quantum resonance condition of FGR. For an anharmonic bath, some deviations corresponding to mismatched quantum-classical resonance condition is possible. We have not explored this correspondence in this work.}



 \mycomment{In the context of Marcus inverted regime, quantum tunneling is considered a major contributor to the rate constants.\cite{siders1981quantum,menzeleev2011direct,lawrence2018analytic,heller2020instanton,richardson2024nonadiabatic,de2024tunneling} In this work, however, AFSSH (with no quantum nuclear effects) gets the rate constants closer to HEOM (or the FGR results) than the Marcus rate constants for the parameters explored. As shown in Eq. \ref{eq_kAFSSH_QCF}, the analytical AFSSH rate is proportional to the FGR rate, but with incorrect quantum correction factor which proves AFSSH can obtain correct trends for the rate constants. The incorrect quantum factor leads to the incorrect long-time population. Hence we conclude that full classical treatement of nuclei can capture the dynamics qualitatively in the deep inverted regime. With this understanding, we propose a careful analysis of nuclear tunneling effects in the deep inverted regime and establish a connection with full classical treatement for future. 
}





\mycomment{Coming to the long-time populations shown in Fig. \ref{rate_population} (d)-(f), AFSSH is unable to capture detailed balance.}  To the best of our knowledge, such a large deviation of AFSSH from the Boltzmann population has not been reported in the literature. The common argument why surface hopping usually obeys detailed balance is:\cite{parandekar2005mixed,subotnik2016understanding,jain2022pedagogical}
\begin{equation}\label{tully_arg}
      \frac{k_b}{k_f} \approx \frac{\int_{v_0}^{\infty}dv \ v.d_{21} Re(c_2^*c_1)}{\int_{0}^{\infty}dv \ v.d_{12} Re(c_1^*c_2)},
\end{equation}
where $v$ is the velocity along the derivative coupling vector and $v_0$ is the minimum velocity required for the transition from a lower energy state to a higher energy state (the frustrated hop condition).
Equation \ref{tully_arg} assumes that the velocities are always in the Boltzmann distribution. Furthermore, assuming $c_1c_2^*$ is independent of both  the velocity and the active surface, it is easy to show that Eq. \ref{tully_arg} gives population ratio to be the Boltzmann answer.

Schmidt, Parandekar, and Tully in 2008 showed that surface hopping obtains detailed balance only approximately.\cite{schmidt2008mixed} They provided a general expression for the rate constants (and therefore long time populations) without assuming self-consistency. Assuming velocity to be at thermal equilibrium, a more general expression for the ratio of rate constants is\cite{tully1990molecular,schmidt2008mixed,granucci2007critical,fang1999improvement}
\begin{equation}\label{true_arg}
    \frac{k_b}{k_f} = \frac{\int_{v_0}^{\infty}dv\frac{ \ v.d_{12} Re(c_\mycomment{2}^*c_\mycomment{1}) }{\vert c_\mycomment{1}\vert^2}h(\dot{|c_\mycomment{2}|^2})} {\int_{0}^{\infty}dv\frac{ \ v.d_{21} Re(c_\mycomment{1}^*c_\mycomment{2}) }{\vert c_\mycomment{2}\vert^2}h(\dot{|c_\mycomment{1}|^2})}.
\end{equation}
We have quantified Eq. \ref{true_arg} with some approximations  for the spin-Boson Hamiltonian in the deep inverted regime (see Eqs. \ref{k_rate_final}-\ref{k_rate_final1}; detailed derivation in SI) and show that surface hopping can significantly deviate from the true Boltzmann result.


A simplistic explanation of why the AFSSH violates detailed balance is as follows. \mycomment{A trajectory initially on adiabatic surface 2 can hop down with no restriction, but has frustrated hop condition for hopping back up. That is, reverse hops from $1\rightarrow2$ will be slower by the Boltzmann factor. Assuming slow decoherence timescale implies net $2\rightarrow1$ transition is faster than the FGR rate constant by the Boltzmann factor. Analogously, a trajectory that starts on surface 1 will have a net $1\rightarrow2$ transition rate slower than the FGR rate by the Boltzmann factor. The net result is that the AFSSH sees the effective temperature as half of the true temperature. The main assumptions made are a slow decoherence timescale and fast vibrational energy relaxation, leading to a large self-consistency issue.}

\begin{figure}[!h]
    \centering
    \includegraphics[width=0.99\linewidth]{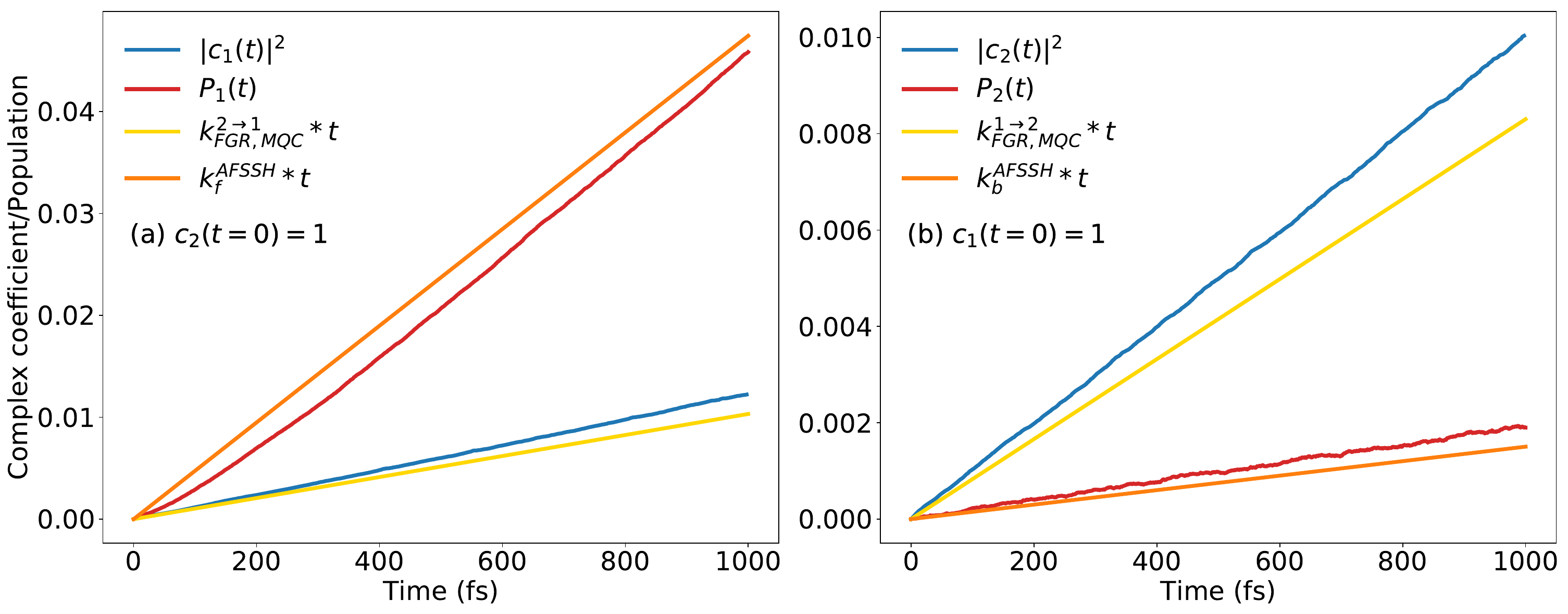}
    \caption{Short time dynamics of  FSSH adiabatic populations and $|c_i|^2$ (in the adiabatic basis) compared with those calculated using FGR rate constants (Eqs. \ref{k_fgr_no_deco} and \ref{k_fgr_reverse}) and AFSSH rate constants (Eqs. \ref{k_rate_final} and \ref{k_rate_final1}). (a) Dynamics initialized on surface \mycomment{2} with $c_\mycomment{2}(t=0)=1$, (b) dynamics initialized on surface \mycomment{1} with $c_\mycomment{1}(t=0)=1$. Parameters are from column two of Table \ref{table_param}.}
    \label{lin_reg}
\end{figure}

We explicitly show the violation of self-consistency of surface hopping (without decoherence) in the adiabatic basis in Fig. \ref{lin_reg} (both populations and $|c_i|^2$ are in adiabatic basis). The results are for the parameters in column 2 of Table  \ref{table_param}. \mycomment{Given the more sensitive nature of this plot, a million trajectories were used with $dt=0.1$ fs.} The two panels show dynamics with different initial conditions: (a) dynamics initialized on surface \mycomment{2} (higher energy state) and (b) initialized on surface \mycomment{1} (lower energy state). MQC FGR rate constant $k_{FGR, MQC}^{i\rightarrow j}t$ matches well with the $|c_j|^2$ -- i.e. the parameters are in the weak adiabatic coupling regime. Importantly, the population calculated using surface hopping (red line in Fig. \ref{lin_reg}) is different from the corresponding $|c_i|^2$. The population of adiabatic surface \mycomment{2}, which is higher in energy, is always less than $|c_\mycomment{2}|^2$ owing to frustrated hops. 

\mycomment{As a curious note, observe that there is an initial consistency in Fig. \ref{lin_reg} (a), but not in Fig. \ref{lin_reg} (b). The inconsistency timescale is dictated by two factors: (a) vibrational energy relaxation, which is roughly $1/\tau=50$ fs, and (b) timescale of sharp fluctuations in quantum amplitudes, which is a few fs. When starting on surface 2, trajectories will hop down on a few fs timescale, and inconsistency will arise on the timescale of vibrational energy relaxation - i.e. roughly 50 fs. When starting from the surface 1, inconsistency will arise within a few fs timescale as trajectories will immediately experience frustrated hops. This is consistent with the observations of Fig. \ref{lin_reg}.
}

\begin{figure}[ht]
    \centering
    \includegraphics[width=0.98\linewidth]{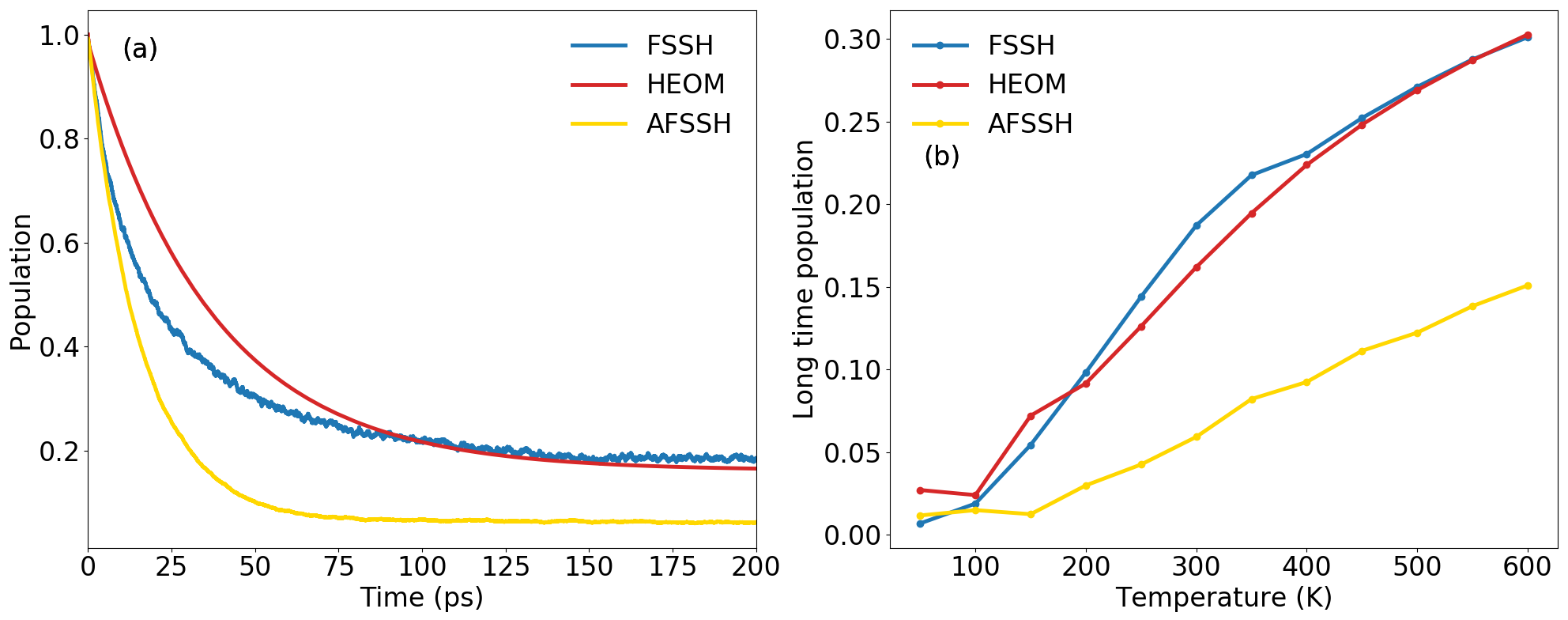}
    \caption{(a) Comparison of diabatic population of state 1 obtained from surface hopping with and without decoherence and HEOM for the parameters in the second column of Table \ref{table_param}. (b) Long-time population comparison between FSSH, AFSSH, and HEOM for a range of  temperatures. Values of other variables are given in table \ref{table_param}, column 2.}
    \label{no_deco}
\end{figure}

We end our discussion with the FSSH results without decoherence in Fig. \ref{no_deco}. The population decay calculated using the  FSSH method is not exponential (Fig. \ref{no_deco} (a)), which is known in literature.\cite{jain2015does} FSSH, however, gets the correct detailed balance (Fig. \ref{no_deco} (b)). This can be understood from Eq. \ref{true_arg}. At thermal equilibrium, for FSSH method (without decoherence) $\langle|c_i|^2\rangle\approx0.5$ and $h(\dot{|c_i|^2})=0.5$ (for $i=1,2$). Therefore, Eq. \ref{true_arg} becomes equivalent to Eq. \ref{tully_arg} and FSSH gives correct detailed balance. 

Although the FSSH dynamics is not exponential, it is still instructive to perform the best fit of an exponential decay to the FSSH population and compute an ad-hoc rate constant. Figure \mycomment{S4} of SI compares the FSSH rate constants so obtained as a function of temperature with those of the analytical theory and AFSSH results (see Fig. \ref{rate_population} (c)). Surprisingly, the FSSH rate constants compare exceptionally well with the analytical rate constants, displaying the erroneous maxima. We do not know the underlying reason for this, given that the analytical rate constants are derived for the linear regime and incorporate decoherence. Since FSSH without decoherence is not recommended because it yields incorrect non-exponential population dynamics, we have not pursued this observation further.



In conclusion, this work investigates the surface-hopping method in the deep Marcus inverted regime. With nearly parallel potential energy surfaces, the decoherence timescale is slow, and the nonadiabatic dynamics in this regime are governed by thermal motion close to the minima of the potential energy surfaces rather than localized dynamics at sharp avoided crossings. Therefore, trajectories can directly drip between adiabatic surfaces without ever accessing regions of strong non-adiabaticity. 

We have developed a model that mimics AFSSH dynamics in the deep inverted regime and derived an analytical formula that predicts the rate constants of the AFSSH method. The derivation effectively provides a rigorous expression for the quantum correction factor for surface hopping simulations. The main assumptions of our model are (a) deep inverted Marcus regime, (b) slow decoherence timescale, and (c) fast vibrational energy relaxation. When these assumptions are valid, we show that the AFSSH rate constants are relatively close to the numerically exact results, but AFSSH works at an effective temperature that is half of the true temperature. 

The AFSSH correctly captures the resonance of the bath modes with the exothermicity of the two-level system and therefore obtains rate constants that are reasonably correct (within a factor of 2). The  inconsistency issue combined with a slow decoherence timescale leads to an incorrect detailed balance. \mycomment{The resonance condition is likely better for a harmonic bath, although an anharmonic bath is not investigated in this work.}


\mycomment{The results of this paper are applicable to systems with nearly parallel energy surfaces with long-time (thermal) dynamics. AFSSH dynamics will be driven by vibrational modes in resonance with the difference in the energy surfaces, and will capture some of the direct transfer between the surfaces away from the crossings. The application of AFSSH to FMO system for example falls in this regime, and its results should be re-investigated from AFSSH perspective. As a final note, this paper shows that AFSSH, with no quantum nuclear effects, captures correct trends for the deep Marcus inverted regime where tunneling is known to be important. This is worth exploring for future work.}





\subsubsection{Acknowledgment}
This paper is supported by the funds received from Science and Engineering Research Board, early research career award Grant no. ECR/2018/001635/CS. PKD thanks the Council of Scientific and Industrial Research (CSIR), India, for his Ph.D. fellowship.  

\bibliography{ref}

\end{document}


\maketitle

\section{Derivation of the \mycomment{mixed quantum-classical} rate equations}

Our aim here is to provide the simplest model that captures the FSSH dynamics. The model is applicable only in the deep inverted Marcus regime and aims to provide qualitative trends for the FSSH results with a constant decoherence timescale.

 We consider the evolution of quantum amplitudes on an adiabatic basis as follows:
\begin{align}
    i\hbar\dot{c_{1}^{ad}}(t) &= V_{1}^{ad}c_{1}^{ad}(t) -i\hbar\Vec{v}\cdot\Vec{d}_{12} c_{2}^{ad}(t),  \\
    i\hbar\dot{c_{2}^{ad}}(t) &= i\hbar\Vec{v}\cdot\Vec{d}_{12}c_{1}^{ad}(t) + V_{2}^{ad}c_{2}^{ad}(t).
\end{align}
$V_i^{ad}$ are the adiabatic potential energies and $c_i^{ad}$ are the adiabatic coefficients, respectively. The initial conditions are $c_1^{ad}(t=0)=1$ for the transition from surfaces $1\rightarrow2$ and $c_2^{ad}(t=0)=1$ for $2\rightarrow1$ transition. 

\subsection{Transition from $\mycomment{2}\rightarrow\mycomment{1}$}

\subsubsection{Fermi's golden rule rate}
We derive the Fermi's golden rule rate in the classical limit (i.e., bath coordinates are treated classically) in the adiabatic basis. This follows the derivation of the Marcus theory, except that the coupling is considered to be derivative coupling.\cite{Nitzan,bhattacharyya2013adiabatic} Importantly, the coupling is seen as a time-dependent external perturbation with the frequencies $\omega_j$ arising due to the velocities.

Defining $\widetilde{c_{j}^{ad}} = c_{j}^{ad}  \exp\left(\frac{i}{\hbar} \int_{0}^{t} V_{j}^{ad}(x(t')) \, dt'\right)$, we obtain
\begin{align}
    \dot{\widetilde{c_{\mycomment{1}}^{ad}}}(t) = \widetilde{c_{\mycomment{2}}^{ad}}(t)\left(\vec{v}\cdot\vec{d}_{12}\right)   \exp\left(\frac{i}{\hbar} \int_{0}^{t} \left(V_{\mycomment{1}}^{ad}-V_{\mycomment{2}}^{ad}\right) \, dt'\right). 
    \label{eq_c2tilde_dot}
\end{align}
For ease of notation, the time dependence of $\vec{v}$, $\Vec{d}_{12}$ and $V_j^{ad}$ is not explicitly written. To solve Eq. \ref{eq_c2tilde_dot}, we assume weak coupling $(\Vec{v}\cdot\vec{d}_{12})$. Therefore, we assume that the adiabatic energy difference is the same as the diabatic energy difference (Eq. 1 of Main paper):
\begin{align}
    \delta V_{ad}^+\equiv V_{\mycomment{1}}^{ad}-V_{\mycomment{2}}^{ad} &\approx \Delta G + \sum_j g_j x_j(t) + \lambda.
    \label{eq_Vad}
\end{align}
with $\lambda=\sum_j g_j^2/(2m\omega_j^2)$. For the Hamiltonian of Eq. 1 of main paper, the $j$'th component of the derivative coupling can be shown to be of the form
\begin{align}
    {d_{j}^{12}} = \frac{V_{c}}{(\delta V_{ad}^+)^{2}+4V_c^2}g_{j} \approx \frac{V_{c}}{(\delta V_{ad}^+)^{2}}g_{j},
    \label{eq_d12}
\end{align}
where we assume $\delta V_{ad}^+ \gg V_c$.

Taking the bath mode velocities to be the same as the diabatic velocities (weak adiabatic coupling assumption) gives
\begin{align}
    {v_{j}}(t) &\approx {v_{j}^{0}} \cos\left(\omega_{j}t\right) - {x_{j}^{0}}\omega_{j}\sin(\omega_{j}t), \\
    \Vec{v}\cdot\Vec{d}_{12} &= \sum_{j}\frac{V_{c}}{(\delta V_{ad}^+)^{2}} g_{j} \left[\left(\frac{v_{j}^{0} - ix_{j}^{0}\omega_{j}}{2}\right)\exp\left(-i\omega_{j}t\right) + \left(\frac{v_{j}^{0} + ix_{j}^{0}\omega_{j}}{2}\right)\exp(i\omega_{j}t)\right],
    \label{eq_vdotd}
\end{align}
where $x_{j}^{0}$ and $v_{j}^{0}$ are the initial positions and velocities, respectively, of the $j$th bath mode.
Substituting Eq. \ref{eq_vdotd} into Eq. \ref{eq_c2tilde_dot}, and making the rotating wave approximation (with $\delta V_{ad}^+<0$), we get,
\begin{align}
    \dot{\widetilde{c_{\mycomment{1}}^{ad}}} \approx \widetilde{c_{\mycomment{2}}^{ad}}\sum_{j}\frac{V_{c}}{(\delta V_{ad}^+)^{2}} g_{j} \left[\left(\frac{v_{j}^{0} + ix_{j}^{0}\omega_{j}}{2}\right)\exp\left(\frac{i}{\hbar}\int_{0}^{t} \left(\delta V_{ad}^+ + \hbar\omega_{j}\right)dt'\right)\right].
    \label{eq_c2tildedot}
\end{align}
We derived the results for a weak nonadiabatic coupling regime. Therefore, for getting a closed form analytical formula, we assume
\(\widetilde{c_{\mycomment{2}}^{ad}}(t) \approx 1\), giving
\begin{align}
    \widetilde{c_{\mycomment{1}}^{ad}}(t) &= \int_{0}^{t}dt_1\sum_{j}\frac{V_{c}}{(\delta V_{ad}^+)^{2}} g_{j} \exp(i\omega_jt_1) \left(\frac{v_{j}^{0} + ix_{j}^{0}\omega_{j}}{2}\right)\exp\left(\frac{i}{\hbar}\int_{0}^{t_1}dt' \delta V_{ad}^+ \right).
    \label{eq_c2ad}
\end{align}
For the simplest analytical expression of the rate constant, we take the explicit dependence of $\delta V_{ad}^+$ on the bath coordinates only in the exponential form in Eq. \ref{eq_c2ad}. $\delta V_{ad}^+$ in the denominator of Eq. \ref{eq_c2ad} originates from the derivative coupling (Eq. \ref{eq_d12}). We for now ignore the explicit bath dependence of $\vec{d}_{12}$ on $\vec{x}$ and use the average $\langle\delta V_{ad}^+\rangle$ in the denominator of Eq. \ref{eq_c2ad} as a simple guess. However, the bath dependence on $\vec{d}_{12}$ causes sharp fluctuations in $|\widetilde{c_{\mycomment{1}}^{ad}}|^{2}$, which are important for capturing the FSSH dynamics and will be addressed later. Using Eq. \ref{eq_Vad} and taking a thermal average we get
\begin{align}      
\label{eq_c2ad2}
\left\langle|\widetilde{c_{\mycomment{1}}^{ad}}|^{2}\right\rangle = \frac{V_{c}^{2}}{4\langle\delta V_{ad}^+\rangle^{4}}\int_{0}^{t}dt_{1}\int_{0}^{t}dt_{2} &\exp\left(-\frac{i}{\hbar}(\Delta G + \lambda)(t_{1} - t_{2})\right)\\
&\sum_{j}\sum_k g_{j} g_k \exp(-i\omega_jt_1) \exp(i\omega_kt_2) S,\notag
\end{align}
where
\begin{align}
S=\left\langle (v_{j}^0 - ix_j^0\omega_j)(v_{k}^0 + ix_k^0\omega_k)  \exp\left(\frac{i}{\hbar}\int_{t_{1}}^{t_{2}}dt'\sum_{l}g_{l}x_{l}(t')\right)\right\rangle.
\end{align}
Using cumulant expansion
\begin{align}
    S\approx&\left\langle (v_j^0)^2+(x_j^0)^2\omega_j^2     \right\rangle\delta_{jk}-\notag\\
    &\frac{1}{2\hbar^2}\sum_l\sum_{l'}g_lg_{l'}\int_{t_{1}}^{t_{2}}dt'\int_{t_{1}}^{t_{2}}dt''\left\langle(v_{j}^0 - ix_j^0\omega_j)(v_{k}^0 + ix_k^0\omega_k)x_l(t')x_{l'}(t'')\right\rangle.
    \label{eq_cumulant}
\end{align}
We assume that bath relaxation is much faster than the population relaxation timescale $t$. Therefore, we neglect terms $\langle v_j^0 x_j(t') \rangle$ and $\langle x_j^0 x_j(t') \rangle$, giving
\begin{align}
    S&=\left\langle (v_j^0)^2+(x_j^0)^2\omega_j^2     \right\rangle\delta_{jk}\left(1-
    \frac{1}{2\hbar^2}\sum_l g_l^2\int_{t_{1}}^{t_{2}}dt'\int_{t_{1}}^{t_{2}}dt''\left\langle x_l(t')x_{l}(t'')\right\rangle\right).
    \label{eq_cumulant1}
\end{align}
Converting the cumulant expansion back to the exponential form and using the classical thermal averages of the position and velocity for the harmonic oscillator yields
\begin{align}
    S=\frac{2k_BT}{m}\delta_{jk}\exp\left(-
    \frac{1}{2\hbar^2}\sum_l g_l^2\int_{t_{1}}^{t_{2}}dt'\int_{t_{1}}^{t_{2}}dt''\left\langle x_l(t')x_{l}(t'')\right\rangle\right).
    \label{eq_cumulant2}
\end{align}
Using $x_j(t)=x_j^0\cos(\omega_j t) + v_j^0/\omega_j\sin(\omega_j t) $ simplifies the time correlation function as
\begin{align}
    \int_{t_{1}}^{t_{2}}dt'\int_{t_{1}}^{t_{2}}dt''\left\langle x_{l}(t')x_{l}(t'') \right\rangle &= \int_{t_{1}}^{t_{2}}dt'\int_{t_{1}}^{t_{2}}dt'' \frac{k_{B}T}{m\omega_{l}^{2}}\cos\left(\omega_{l}(t' - t'')\right),\\
    &=\frac{2k_{B}T}{m\omega_{l}^{4}}\left(1-\cos(\omega_l(t_1-t_2))\right).
    \label{eq_xcorr}
\end{align}
Substituting Eq. \ref{eq_xcorr} in Eq. \ref{eq_cumulant2} gives
\begin{align}
    S=\frac{2k_BT}{m}\delta_{jk}\exp\left(-
    \frac{k_{B}T}{\hbar^2}\sum_l  \frac{g_l^2}{m\omega_{l}^{4}}\left(1-\cos(\omega_l(t_1-t_2))\right) \right).
    \label{eq_cumulant3}
\end{align}
We further simplify Eq. \ref{eq_cumulant3} in the short $t_1-t_2$ limit:
\begin{align}
    S&=\frac{2k_BT}{m}\delta_{jk}\exp\left(-
    \frac{k_{B}T}{\hbar^2}\sum_l  \frac{g_l^2}{2m\omega_{l}^{2}}(t_1-t_2)^2) \right),\\
    &=\frac{2k_BT}{m}\delta_{jk}\exp\left(-
    \frac{k_{B}T\lambda}{\hbar^2}(t_1-t_2)^2) \right).
    \label{eq_cumulant4}
\end{align}

Finally, substituting Eq. \ref{eq_cumulant4} back in Eq. \ref{eq_c2ad2} gives
\begin{align}      \left\langle|\widetilde{c_{\mycomment{1}}^{ad}}|^{2}\right\rangle = \frac{V_{c}^{2}k_BT}{2\langle\delta V_{ad}^+\rangle^{4}}\int_{0}^{t}dt_{1}\int_{0}^{t}dt_{2} &\sum_{j} \frac{g_j^2}{m}\exp\left(-\frac{i}{\hbar}(\Delta G + \lambda+\hbar\omega_j)(t_{1} - t_{2})\right)\\
&  \exp\left(-
    \frac{k_{B}T\lambda}{\hbar^2}(t_1-t_2)^2) \right).
\label{eq_c2ad3}
\end{align}

A simple variable transformation $P=t_1-t_2, Q=t_1+t_2$ yields

\begin{align}    \left\langle|\widetilde{c_{\mycomment{1}}^{ad}}|^{2}\right\rangle = \frac{V_{c}^{2}k_BT}{2\langle\delta V_{ad}^+\rangle^{4}}\frac{1}{2}\int_{0}^{2t}dQ\int_{-Q}^{Q}dP \sum_{j} \frac{g_{j}^{2}}{m}\exp\left(-\frac{i}{\hbar}(\Delta G + \lambda + \hbar\omega_{j})P\right)
    \notag\\
    \exp\left(-\frac{\lambda k_{B}TP^{2}}{\hbar^{2}} \right).
    \label{eq_c2_PQ}
\end{align}
 The rate constant $k_{FGR}^{\mycomment{2}\rightarrow \mycomment{1}}$ is calculated as the time derivative of Eq. \ref{eq_c2_PQ} at large $t\rightarrow\infty$. This gives 
\begin{align}
    \boxed{k_{FGR,MQC}^{\mycomment{2}\rightarrow \mycomment{1}} = \sum_{j} \frac{V_{c}^{2}}{2\langle\delta V_{ad}^+\rangle^{4}}\frac{k_{B}T}{m} g_{j}^{2}\sqrt{\frac{\pi\hbar^{2}}{\lambda k_{B}T}} \exp\left(\frac{-(\Delta G + \lambda + \hbar\omega_{j})^{2}}{4\lambda k_{B}T}\right)}
    \label{eq_kFGR_12}
\end{align}
Equation \ref{eq_kFGR_12} is the Fermi's golden rule rate for the transition from adiabat \mycomment{2} to \mycomment{1} in the classical bath limit. This rate constant is the sum of the independent Marcus rates obtained for each bath mode with an exothermicity of $\Delta G+\hbar \omega_j$ and coupling squared of $\hbar^2/2 \langle v^2\rangle \langle d_j^{12}\rangle^2=\hbar^2/2(k_BT/m)(g_j^2V_c^2/\langle\delta V_{ad}^+\rangle^4)$. We take $\langle\delta V_{ad}^+\rangle=\Delta G+\lambda$.


\subsubsection{Modeling FSSH rates}
If the FSSH simulation obeys self-consistency (i.e., $P_\mycomment{1}=|c_\mycomment{1}^{ad}|^2$), then the surface-hopping results will be close to the $k_{FGR}^{\mycomment{2}\rightarrow \mycomment{1}}$ rate constant. However, because of frustrated hops and decoherence, self-consistency is not maintained, and the FSSH rate constant deviates from the $k_{FGR}^{\mycomment{2}\rightarrow \mycomment{1}}$ expression.

\begin{figure}[h]
    \centering
    \includegraphics[width=0.75\linewidth]{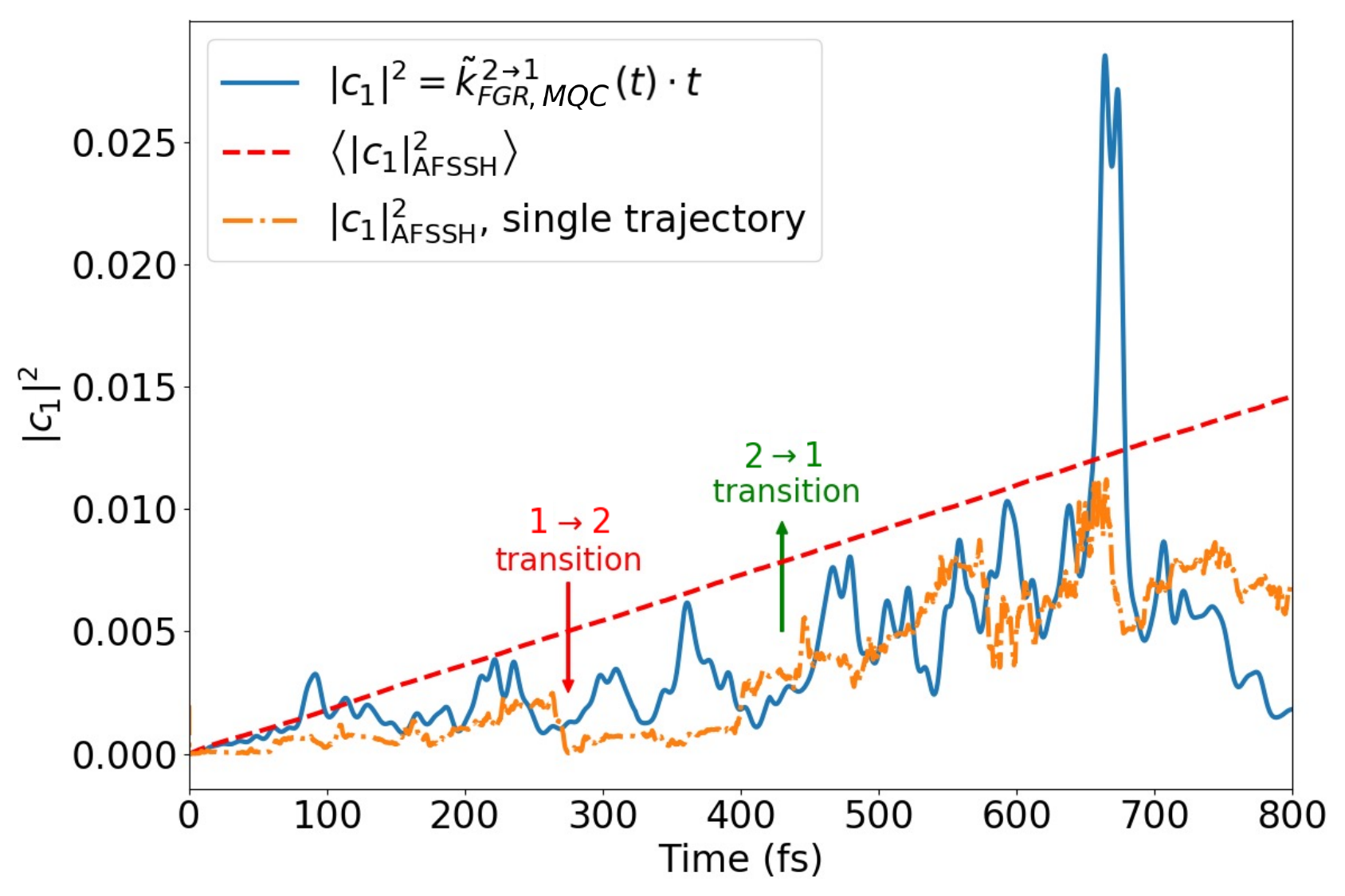}
    \caption{Plot of $|c_\mycomment{1}^2|$ as a function of time for a random trajectory. $|c_\mycomment{1}^2|$ calculated from a FSSH trajectory is compared with $|c_\mycomment{1}^2|$ calculated using Eq. \ref{eq_c2_t}. Averaged $|c_\mycomment{1}^2|$ calculated using FSSH is also shown for reference.}
    \label{fig_c2_1traj}
\end{figure}

To derive a rate constant for the FSSH algorithm, we must model the fluctuations in $|c_\mycomment{1}^{ad}|^2$. Consider a trajectory initialized on adiabatic surface \mycomment{2}. $|c_\mycomment{1}^{ad}|^2$ of a typical trajectory is shown by the orange line in Fig. \ref{fig_c2_1traj}. Although the average $\langle|c_\mycomment{1}^{ad}|^2\rangle$ is linear and has a slope close to the FGR rate constant, $|c_\mycomment{1}^{ad}|^2$ for an individual trajectory fluctuates. A necessary condition for a hop from surfaces \mycomment{2} to \mycomment{1} is $\dot{|c_\mycomment{1}^{ad}|^2}>0$; for example, the green arrow in Fig. \ref{fig_c2_1traj}. A trajectory that hops down to surface \mycomment{1} can hop back to surface \mycomment{2} only when $\dot{|c_\mycomment{1}^{ad}|^2}<0$ and $0.5mv^2>\delta V_{ad}^+$ (frustrated hop condition).

To obtain an analytical expression for the FSSH rates, we first focus on trajectories with initial conditions on surface \mycomment{2} and build a simple kinetic model:
\begin{align}
    \dot{P}_{\mycomment{1}}^{c_\mycomment{2}=1}(t)=- k^{\mycomment{1}\rightarrow \mycomment{2}}_{c_\mycomment{2}=1}P_{\mycomment{1}}^{c_\mycomment{2}=1}+k^{\mycomment{2}\rightarrow \mycomment{1}}_{c_\mycomment{2}=1}P_{\mycomment{2}}^{c_\mycomment{2}=1}, 
    \label{eq_kinetic}
\end{align}
where $P_{i}^{c_1=1}$ is the fraction of trajectories on surface $i$ with $c_1\approx 1$ and $k^{i\rightarrow j}_{c_1=1}$ is the rate constant of hopping from surface $i$ to $j$ with $c_1\approx 1$. We assume that the rate constants $k^{1\rightarrow2}_{c_\mycomment{2}=1}$ and $k^{2\rightarrow1}_{c_\mycomment{2}=1}$ are independent of each other. The thermally averaged FSSH hopping rate constants are given by
\begin{align}
    k_{c_{\mycomment{2}} = 1}^{\mycomment{2}\rightarrow \mycomment{1}} &= \left\langle \frac{\dot{|c_{\mycomment{1}}|^{2}} h(\dot{|c_{\mycomment{1}}|^{2}})}{|c_{\mycomment{2}}|^{2}}\right\rangle,
    \label{eq_k_u2d}\\
    k_{c_{\mycomment{2}} = 1}^{\mycomment{1}\rightarrow \mycomment{2}} &= \left\langle \frac{-\dot{|c_{\mycomment{1}}|^{2}} h(-\dot{|c_{\mycomment{1}}|^{2}})h(0.5mv^2-\delta V_{ad}^+)}{|c_{\mycomment{1}}|^{2}}\right\rangle.
    \label{eq_k_d2u}
\end{align}
$h(x)$ is the Heaviside function, and Eq. \ref{eq_k_d2u} includes the frustrated hop condition. $v$ is the velocity along the reaction coordinate $\hat{d}_{12}$.




To further simplify, we consider the following ansatz for a single FSSH trajectory within the linear time regime in population decay:
\begin{align}
    |c_{\mycomment{1}}|^{2}(t) = \tilde{k}_{FGR,MQC}^{\mycomment{2}\rightarrow \mycomment{1}}(t) t,
    \label{eq_c2_t}
\end{align}
with 
\begin{align}
    \tilde{k}_{FGR,MQC}^{\mycomment{2}\rightarrow \mycomment{1}}(t) = \sum_{j} \frac{V_{c}^{2}}{2(\delta V_{ad}^+(t))^{4}}\frac{k_{B}T}{m} g_{j}^{2}\sqrt{\frac{\pi\hbar^{2}}{\lambda k_{B}T}} \exp\left(\frac{-(\Delta G + \lambda + \hbar\omega_{j})^{2}}{4\lambda k_{B}T}\right).
    \label{eq_kfgr_traj}
\end{align}
$\tilde{k}_{FGR,MQC}^{\mycomment{2}\rightarrow \mycomment{1}}(t)$ is the FGR rate constant with the bath modes explicitly present in $\delta V_{ad}^+$ in the denominator. The intuition behind this simple ansatz is that the thermal bath leads to rapid fluctuations in $\delta V_{ad}^+$, which do not affect the overall average. However, rapid oscillations are critical for modeling FSSH rates. The blue line in Fig. \ref{fig_c2_1traj} shows the results of Eq. \ref{eq_c2_t} for a sample trajectory. Clearly, our ansatz is a simple one that captures the surface hopping results only qualitatively.

Considering this, we obtain the rate as
\begin{align}
    \dot{|c_{\mycomment{1}}|^{2}} &= \tilde{k}_{FGR,MQC}^{\mycomment{2}\rightarrow \mycomment{1}}\left(1 - \frac{4\dot{\delta V_{ad}^+}}{\delta V_{ad}^+} t\right),\\
    &\approx  -\tilde{k}_{FGR,MQC}^{\mycomment{2}\rightarrow \mycomment{1}}\frac{4\dot{\delta V_{ad}^+}}{\delta V_{ad}^+} t.
    \label{eq_c2dot}
\end{align}
Transforming to the reaction coordinate yields $\delta V_{ad}^+=\Delta G+\lambda+gx$, where $x$ is the reaction coordinate, $g = \sqrt{2m\omega_0^{2}\lambda}$, and $\omega_0$ is the reaction coordinate frequency. Therefore $h(\dot{|c_{\mycomment{1}}|^{2}})=h(-v)$. Putting back in Eq. \ref{eq_k_u2d} gives
\begin{align}
    k^{\mycomment{2}\rightarrow \mycomment{1}}_{c_{\mycomment{2}} = 1} &\approx k_{FGR,MQC}^{\mycomment{2}\rightarrow \mycomment{1}} \frac{4g\left\langle|v|\right\rangle}{\langle\delta V_{ad}^+\rangle}t=\dot{k}^{\mycomment{2}\rightarrow \mycomment{1}}_{c_{\mycomment{2}} = 1}t,
    \label{eq_k_u2d2}
\end{align}
with $\dot{k}_{c_{\mycomment{2}}=1}^{2\rightarrow1}=k_{FGR,MQC}^{\mycomment{2}\rightarrow \mycomment{1}} \frac{4g\left\langle|v|\right\rangle}{\langle\delta V_{ad}^+\rangle}$. We assumed $|c_\mycomment{2}|^2\approx 1$ and took independent averages to obtain Eq. \ref{eq_k_u2d2}.
$\langle|v|\rangle=\sqrt{\frac{k_B T}{2\pi m}}$ is the average thermal speed. Furthermore, note that $k^{\mycomment{2}\rightarrow \mycomment{1}}_{c_{\mycomment{2}} = 1}$ is a linear function of time $t$. 

Substituting Eqs. \ref{eq_c2_t} and \ref{eq_c2dot} in Eq. \ref{eq_k_d2u} gives
\begin{align}
    k^{\mycomment{1}\rightarrow \mycomment{2}}_{c_{\mycomment{2}} = 1} =  \frac{4g\left\langle|v|h(v-v_0) \right\rangle}{\langle\delta V_{ad}^+\rangle}\approx\frac{4g\left\langle|v|\right\rangle}{\langle\delta V_{ad}^+\rangle}e^{-\beta \Delta G},
    \label{eq_k_d2u2}
\end{align}
where $0.5mv_0^2=\langle \delta V_{ad}^+\rangle$. For simplicity, we consider that $\langle \delta V_{ad}^+\rangle\approx \Delta G$.

Substituting Eqs. \ref{eq_k_u2d2} and \ref{eq_k_d2u2} in Eq. \ref{eq_kinetic}
\begin{align}
    \dot{P}_{\mycomment{1}}^{c_\mycomment{2}=1}(t) &= - (k_{c_{\mycomment{2}}=1}^{\mycomment{1}\rightarrow \mycomment{2}} + \dot{k}_{c_{\mycomment{2}}=1}^{1\rightarrow2}t)P_{\mycomment{1}}^{c_\mycomment{2}=1} + \dot{k}_{c_{\mycomment{2}}=1}^{2\rightarrow1}t,\\
    &\approx - k_{c_{\mycomment{2}}=1}^{\mycomment{1}\rightarrow \mycomment{2}} P_{\mycomment{1}}^{c_1=1} + \dot{k}_{c_{\mycomment{2}}=1}^{2\rightarrow1}t.
    \label{eq_p2dot}
\end{align}
 Over the time scale of decoherence, we assume $k_{c_{\mycomment{2}}=1}^{\mycomment{1}\rightarrow \mycomment{2}}\gg \dot{k}_{c_{\mycomment{2}}=1}^{\mycomment{2}\rightarrow \mycomment{1}}t$. Solving Eq. \ref{eq_p2dot} gives
\begin{align}
    \label{eq_P2_t}
    P_{\mycomment{1}}^{c_\mycomment{2}=1}(t) =& \frac{\dot{k}_{c_{\mycomment{2}}=1}^{\mycomment{2}\rightarrow \mycomment{1}}}{k_{c_{\mycomment{2}}=1}^{\mycomment{1}\rightarrow \mycomment{2}}}t + \frac{\dot{k}_{c_{\mycomment{2}}=1}^{\mycomment{2}\rightarrow \mycomment{1}}}{(k_{c_{\mycomment{2}}=1}^{\mycomment{1}\rightarrow \mycomment{2}})^2}[\exp(-k_{c_{\mycomment{2}}=1}^{\mycomment{1}\rightarrow \mycomment{2}}t) - 1], \\
    \dot{P}_{\mycomment{1}}^{c_\mycomment{2}=1}(t) =& \frac{\dot{k}_{c_{\mycomment{2}}=1}^{\mycomment{2}\rightarrow \mycomment{1}}}{k_{c_{\mycomment{2}}=1}^{\mycomment{1}\rightarrow \mycomment{2}}}\left[1- \exp(-k_{c_{\mycomment{2}}=1}^{\mycomment{1}\rightarrow \mycomment{2}}t) \right].
    \label{eq_P2dot2}
\end{align}

\begin{figure}[h]
    \centering
    \includegraphics[width=0.75\linewidth]{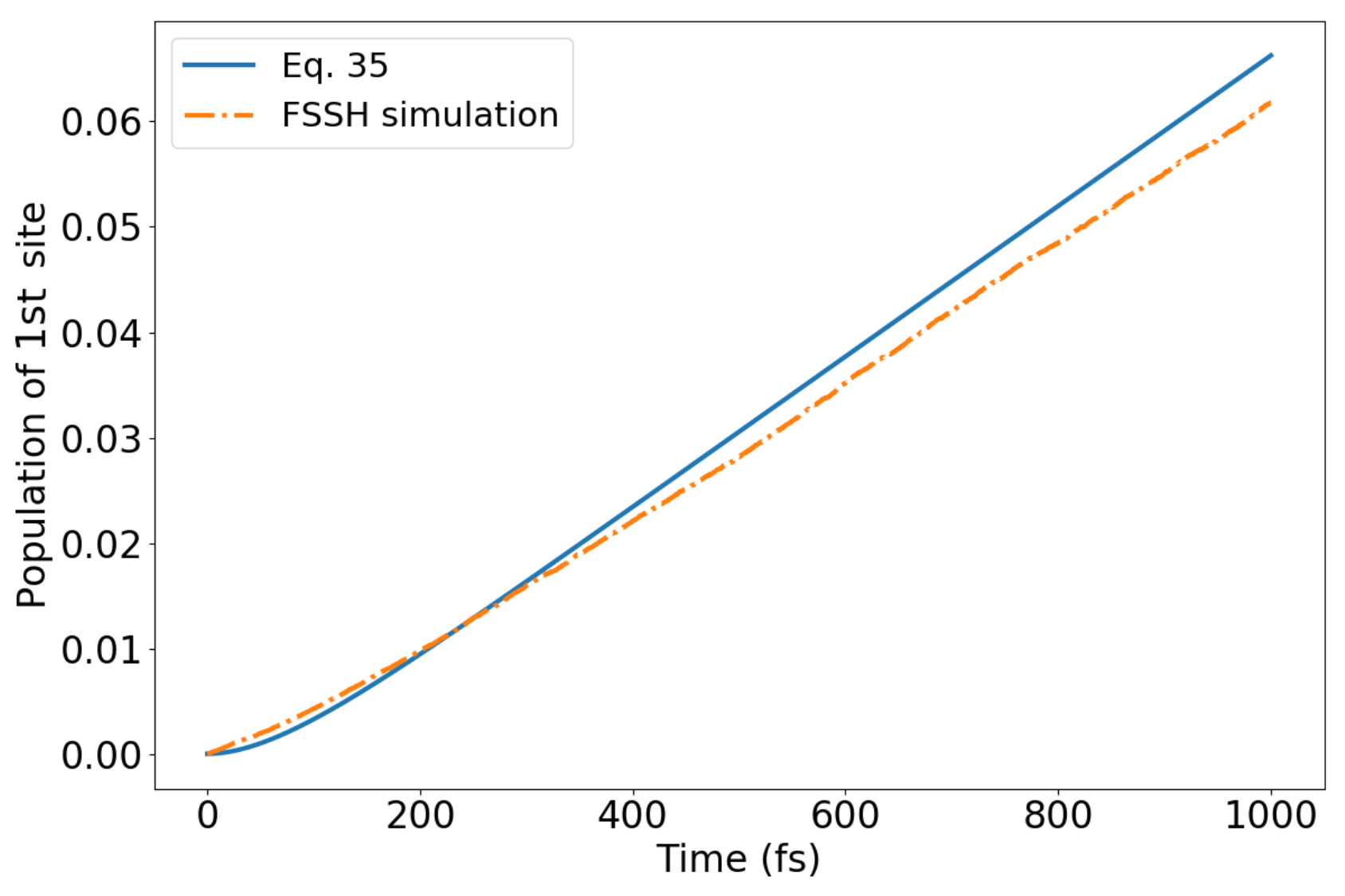}
    \caption{Comparison of population of adiabat \mycomment{1}, with initial conditions on surface \mycomment{2} calculated using FSSH simulation (no decoherence) and using Eq. \ref{eq_P2_t}.}
    \label{fig_eq_35}
\end{figure}

The numerical evidence for Eq. \ref{eq_P2_t} is shown in Fig. \ref{fig_eq_35} which compares results from Eq. \ref{eq_P2_t} with those calculated numerically from FSSH simulations without decoherence for the parameters of column 2 of Table 1 of the Main paper. Note the good agreement up to 1 ps between the two. Furthermore, the FSSH results also show nonlinearity at very early time scales ($\approx$ 20 fs), which matches our analytical predictions of Eqs. \ref{eq_P2_t} and \ref{eq_P2dot2}.

The rate constant in Eq. \ref{eq_P2dot2} is not a constant in time. We average Eq. \ref{eq_P2dot2} over the decoherence time scale. We consider the simplest decoherence model: a constant collapse timescale $\tau$. That is, the probability of a collapse is $dt/\tau$, where $dt$ is a small time-step. This gives the probability distribution that a collapse will occur at time $t$ to be:
\begin{align}
    \rho (t) =& \frac{\exp{(-t/\tau)}}{\int_{0}^{\infty}dt \exp(-t/\tau)},\\
    =& \frac{1}{\tau}\exp(-t/\tau).
    \label{eq_rho}
\end{align}
Finally, averaging Eq. \ref{eq_P2dot2} over the distribution $\rho(t)$ gives the AFSSH rate constant for $\mycomment{2}\rightarrow \mycomment{1}$ transition:
\begin{align}
     k_{f}^{AFSSH} =  \int_0^\infty dt \rho(t)  \dot{P}_{\mycomment{1}}^{c_\mycomment{2}=1}(t)= \frac{\dot{k}_{c_{\mycomment{1}}=1}^{\mycomment{2}\rightarrow \mycomment{1}}}{k_{c_{\mycomment{2}}=1}^{\mycomment{1}\rightarrow \mycomment{2}}}\left(\frac{k_{c_{\mycomment{2}}=1}^{\mycomment{1}\rightarrow \mycomment{2}}\tau}{1+k_{c_{\mycomment{2}}=1}^{\mycomment{1}\rightarrow \mycomment{2}}\tau}\right).
\end{align}
Using Eq. \ref{eq_k_u2d2} and \ref{eq_k_d2u2} gives the final AFSSH rate constant:
\begin{align}
    \boxed{k_f^{AFSSH}=k_{FGR,MQC}^{\mycomment{2}\rightarrow \mycomment{1}}  \frac{k_{c_{\mycomment{2}}=1}^{\mycomment{1}\rightarrow \mycomment{2}} \tau }{1+k_{c_{\mycomment{2}}=1}^{\mycomment{1}\rightarrow \mycomment{2}} \tau } e^{\beta \Delta G}}
\end{align}

\subsection{Transition from $\mycomment{1}\rightarrow \mycomment{2}$}
Following arguments similar to the $\mycomment{2}\rightarrow \mycomment{1}$ transition with the initial condition $c_\mycomment{1}(t=0)=1$, Fermi's golden rule rate derivation remains the same, except $\Delta G$ switches to $-\Delta G$ and the resonance condition in Eq. \ref{eq_c2tildedot} changes $\hbar\omega_j$ to $-\hbar\omega_j$. Therefore,
\begin{align}
    \boxed{k_{FGR,MQC}^{\mycomment{1}\rightarrow \mycomment{2}} = \sum_{j} \frac{V_{c}^{2}}{2\langle\delta V_{ad}^-\rangle^{4}}\frac{k_{B}T}{m}g_{j}^{2} \sqrt{\frac{\pi\hbar^{2}}{\lambda k_{B}T}} \exp\left(\frac{-(\Delta G - \lambda + \hbar\omega_{j})^{2}}{4\lambda k_{B}T}\right)}
\end{align}
where $\langle\delta V_{ad}^-\rangle=\Delta G-\lambda$.

The FSSH hopping rates follow a similar derivation with $c_\mycomment{1}\approx1$, giving analogous expressions:
\begin{align}
    k^{\mycomment{1}\rightarrow\mycomment{2}}_{c_{\mycomment{1}} = 1} &\approx k_{FGR,MQC}^{\mycomment{1}\rightarrow \mycomment{2}} \frac{4g\left\langle|v|\right\rangle}{\langle\delta V_{ad}^-\rangle}e^{-\beta \Delta G}t=\dot{k}^{\mycomment{1}\rightarrow \mycomment{2}}_{c_{\mycomment{1}} = 1}t,\\
    k^{\mycomment{2}\rightarrow \mycomment{1}}_{c_{\mycomment{1}} = 1}& \approx\frac{4g\left\langle|v|\right\rangle}{\langle\delta V_{ad}^-\rangle}.
\end{align}

This results in a net backward AFSSH rate constant of
\begin{align}
    \boxed{k_{b}^{AFSSH} = k_{FGR,MQC}^{\mycomment{1}\rightarrow \mycomment{2}}{\left(\frac{k^{\mycomment{2}\rightarrow \mycomment{1}}_{c_{\mycomment{1}}=1}\tau}{1+ k^{ \mycomment{2}\rightarrow \mycomment{1}}_{c_{\mycomment{1}}=1}\tau}\right)}e^{-\beta \Delta G}}
\end{align}

We make one final note about decoherence corrections in our analytical formula. Consider a trajectory initialized on surface \mycomment{2} (with $c_\mycomment{2}(t=0)=1$) that hops to surface \mycomment{1}. A decoherence event on surface \mycomment{1} will give $c_\mycomment{1}=1$. Therefore, decoherence provides a transfer rate between the $1\rightarrow 2$ transitions and $2\rightarrow 1$ transitions. We have not included these transitions in our model, since including them leads to four coupled differential equations, which are not solvable to provide an analytical rate constant.

\section{Derivation of the FGR rate constant}
We assume weak diabatic coupling $V_c$ and weak reorganization energy $\lambda$ in this work. It is worth deriving the FGR rate constants in this regime from two approaches to gain better insights. First, we 
start with the weak $V_c$ limit and work in the more common diabatic basis. Second, we show the FGR derivation in the adiabatic basis assuming weak non-adiabatic coupling.

\subsection{Diabatic basis}
In the limit of weak diabatic coupling, the FGR rate constant in the diabatic basis is given by
\begin{align}
    k_{FGR}^{d} = \frac{\vert V_c\vert ^2}{\hbar}\int_{-\infty}^{\infty} dt \ e^{i \Delta G t/\hbar} \textrm{exp} \left\{ -\frac{1}{\hbar}\int_0^\infty d\omega \frac{J(\omega)}{\pi\omega^2} \left(\textrm{coth}\left( \frac{\beta \hbar \omega}{2}\right) (1-\cos(\omega t) ) - i \sin\omega t \right) \right\}.
    \label{eq_k_FGR}
\end{align}

We discussed in the main paper that the small time approximation is not valid for the parameters investigated in this work. Instead we make the weak $\lambda$ assumption. Taking $J(\omega)\propto\lambda$, we Taylor expand the term containing $J(\omega)$ to first order. This gives
\begin{align}
    k_{FGR}^{d} \approx \frac{\vert V_c\vert ^2}{\hbar}\int_{-\infty}^{\infty} dt \ e^{i \Delta G t/\hbar} \left(1-\left\{ \frac{1}{\hbar}\int_0^\infty d\omega \frac{J(\omega)}{\pi\omega^2} \left(\textrm{coth}\left( \frac{\beta \hbar \omega}{2}\right) (1-\cos(\omega t) ) - i \sin\omega t \right) \right\}\right).
    \label{eq_k_FGR_Taylor}
\end{align}

Using $\int_{-\infty}^\infty dt e^{iEt/\hbar}=2\pi\delta(E/\hbar)$, Eq. \ref{eq_k_FGR_Taylor} can be easily simplified to (assuming $\Delta G\ne 0$) 
\begin{align}
k_{FGR}^{d}
&= \frac{|V_c|^2}{\hbar}    \int_0^\infty d\omega \frac{J(\omega)}{\hbar \pi\omega^2}
   \left[
   \begin{aligned}
      \coth\!\left(\frac{\beta\hbar\omega}{2}\right)
      \pi\!\left(
        \delta(\Delta G/\hbar-\omega)
        +\delta(\Delta G/\hbar+\omega)
      \right)
      \\
      +\pi\!\left(
        \delta(\Delta G/\hbar+\omega)
        -\delta(\Delta G/\hbar-\omega)
      \right)
   \end{aligned}
   \right].
\label{eq_k_FGR_dirac}
\end{align}
Simplifying, we get rate constant expressions for $\Delta G<0$ (i.e. $2\rightarrow 1$ transition) and $\Delta G>0$ (i.e. $1\rightarrow 2$ transition):
\begin{align}
\label{eq_FGR_ex_12}
k_{FGR}^{2\rightarrow 1} &= \frac{2k_BTV_c^2}{\hbar |\Delta G|^3}J\left(\frac{|\Delta G|}{\hbar}\right) \frac{\beta|\Delta G|}{1-e^{-\beta|\Delta G|}},\\
k_{FGR}^{1\rightarrow 2} &= \frac{2k_BTV_c^2}{\hbar |\Delta G|^2}J\left(\frac{|\Delta G|}{\hbar}\right) \frac{\beta|\Delta G|}{1-e^{-\beta|\Delta G|}}e^{-\beta|\Delta G|}.
\label{eq_FGR_ex_21}
\end{align}

We have multiplied and divided by $\beta|\Delta G|$ in Eqs. \ref{eq_FGR_ex_12} and \ref{eq_FGR_ex_21} to come to the familiar harmonic quantum correction factor. This will be useful in comparing the AFSSH rate constants which gets this quantum correction factor wrong. See the main text for a detailed discussion.

\subsection{Adiabatic basis}
FGR rate constant has been derived in the adiabatic basis treating non-adiabatic coupling as the perturbation. For completeness, we provide a brief derivation here as well.

In the adiabatic basis, the coupling between the two electronic states is given by $\sum_j i\hbar \hat{p}_jd_{j}^{12}/m$ where the summation index $j$ is over the bath modes, $\hat{p}_j$ is the momentum operator for the $j$th bath mode and $d_j^{12}$ is the corresponding derivative coupling. In the weak $V_c$ limit, we take the $d_j^{12}\approx\frac{V_{c}}{(\Delta G)^{2}}g_{j}$ as a constant.

The thermally FGR rate constant can be written as
\begin{align}
    k_{FGR}^{ad} = \frac{2\pi}{\hbar} \sum_I\sum_J \frac{e^{-\beta E_I}}{Q} \left|\left\langle I| \sum_j\frac{i\hbar \hat{p}_jd_{j}^{12}}{m}| J\right\rangle \right|^2 \delta (\Delta G + E_J-E_I),
    \label{eq_k_FGR_ad}
\end{align}
where $I$ ($J$) is the summation index for bath modes for the initial (final) state, and $E_I$ ($E_J$) is the corresponding bath eigen energy.

$|I\rangle$ and $|J\rangle$ are eigen functions of shifted harmonic oscillators. We denote $j_k^I$ as the $k$th eigen function of the $j$th bath mode for the initial state $I$, and correspondingly define $j_{k\prime}^J$. 
In the limit of weak reorganization energy $\lambda\rightarrow 0$, we take 
\begin{align}
    \langle j_k^I|\hat{p}_j|j_{k\prime}^J\rangle=i\sqrt{\frac{\hbar m \omega_j}{2}} \left[\sqrt{j_k}\delta_{j_k^I,j_{k\prime}^J-1}  - \sqrt{j_k+1}\delta_{j_k^I,j_{k\prime}^J+1}   \right].
\end{align}
Importantly, we have assumed $\langle j_k^I | j_k^J\rangle\approx 1$ in the weak $\lambda$ limit. With this assumption, Eq. \ref{eq_k_FGR_ad} becomes
\begin{align}
     k_{FGR}^{ad} = \frac{2\pi}{\hbar} \sum_j \left(\frac{V_cg_j}{\Delta G^2}\right)^2 \frac{\sum_{j_k^I}e^{-\beta \hbar \omega_j j_k^I} \frac{\hbar^3 \omega_j}{2m}\left[(j_k^I+1)\delta(\Delta G+\hbar\omega_j) + j_k^I\delta(\Delta G-\hbar\omega_j)  \right]     }
     {\sum_{j_k^I}e^{-\beta \hbar \omega_j j_k^I}}.
\end{align}
The sum of $j_k^I$ can be easily evaluated to give

\begin{align}
     k_{FGR}^{ad} = \frac{2\pi}{\hbar} \sum_j \left(\frac{V_cg_j}{\Delta G^2}\right)^2  \frac{\hbar^2 \omega_j}{2m}\frac{1}{1-e^{-\beta \hbar \omega_j}}\left[\delta(\Delta G/\hbar+\omega_j) + e^{-\beta \hbar \omega_j}\delta(\Delta G/\hbar-\omega_j)\right] .\label{eq_FGR_ad_delta}
\end{align}

Taking $J(\omega)= \frac{\pi}{2} \sum_{j} \frac{g_j^2}{m\omega_j}\delta(\omega-\omega_j)$ and the convention $\Delta G<0$ as $2\rightarrow 1$ transition and $\Delta G>0$ as $1\rightarrow 2$ transition, Eq. \ref{eq_FGR_ad_delta} simplifies to give Eqs. \ref{eq_FGR_ex_12} and \ref{eq_FGR_ex_21}.

\subsection{Small time approximation in diabatic basis}
We come back to the FGR in the diabatic basis given by Eq. \ref{eq_k_FGR}. High temperature and small time approximation to this equation gives Marcus rate constant. We show numerically here for one example that though the high temperature approximation is valid, small time approximation is not valid. 

\begin{figure}[h!]
    \centering
    \includegraphics[width=0.6\linewidth]{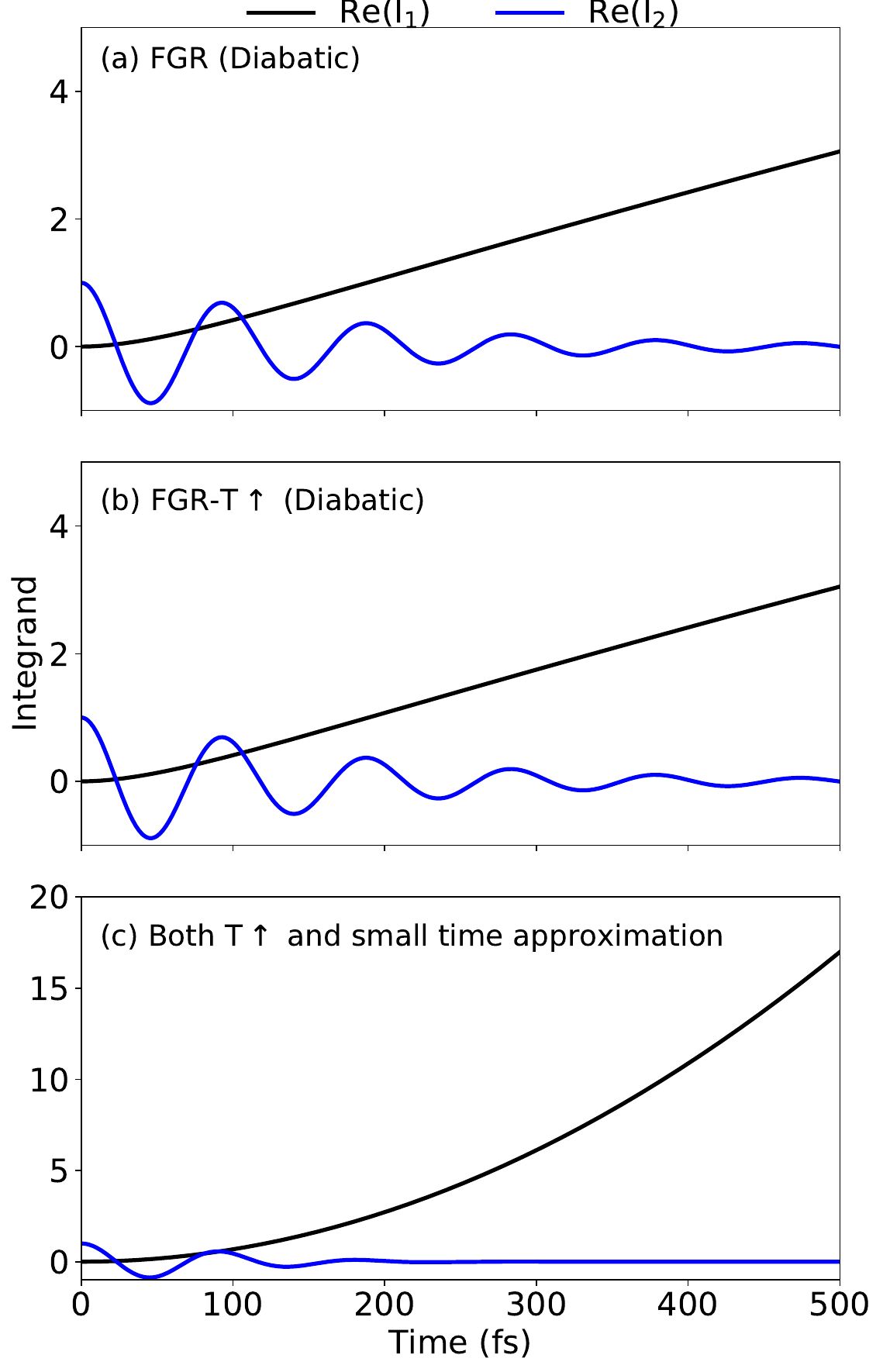}
    \caption{Plots of $Re(I_1)$ and $Re(I_2)$ defined in the text, with the exact expression shown in (a), high temperature approximation shown in (b) and high temperature and small time approximation shown in (c). The parameters are for column 2 of Table 1 of the main paper.  }
    \label{fig_I1_I2}
\end{figure}

Defining
\begin{align}
    I_1(t)&=\frac{1}{\hbar}\int_0^\infty d\omega \frac{J(\omega)}{\pi\omega^2} \left(\textrm{coth}\left( \frac{\beta \hbar \omega}{2}\right) (1-\cos(\omega t) ) - i \sin\omega t \right),\\
    I_2(t)&=e^{i\Delta Gt/\hbar} e^{-I_1(t)},
\end{align}

we plot $Re(I_1)$ and $Re(I_2)$ as a function of time in Fig. \ref{fig_I1_I2} (a). The high temperature approximation ($\tanh(0.5\beta\hbar\omega)\approx0.5\beta\hbar\omega$) plot is shown in Fig. \ref{fig_I1_I2} (b), and small time approximation in \ref{fig_I1_I2} (c). The plots are made for the standard parameters shown in 2nd column of Table 1 of the main paper. $I_2$ decay in about 300 fs timescale. Over this timescale, high temperature approximation is valid, but there is a large deviation when small time approximation is made.

\section{Rate constants from FSSH simulations}

\begin{figure}[ht]
    \centering
    \includegraphics[width=0.75\linewidth]{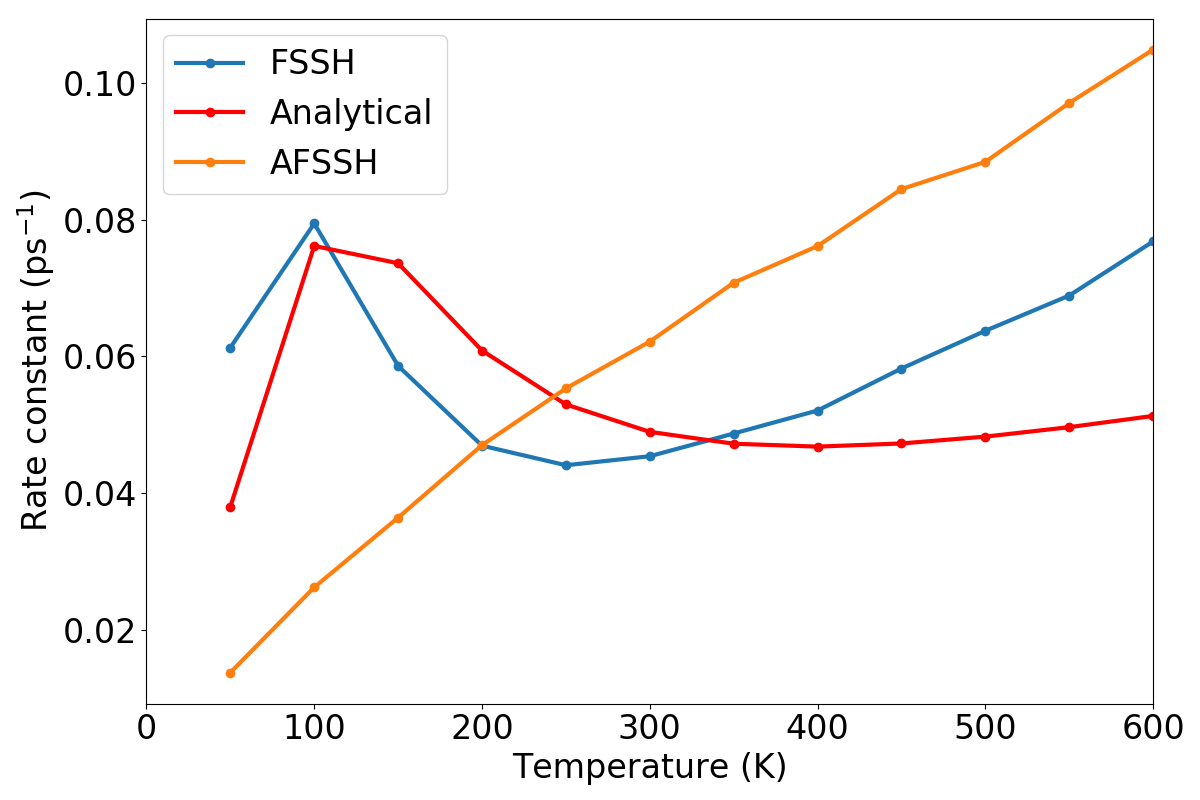}
    \caption{Rate constants obtained from FSSH, AFSSH, and analytical simulations as a function of temperature. The FSSH simulations show non-exponential dynamics, and the rate constants represent the best fit of an exponential function to the population dynamics. There is a surprising match between the analytical and the FSSH rate constants. }
    \label{fig_FSSH_rates}
\end{figure}

FSSH simulations without decoherence do not yield exponential population decay. Nonetheless, we perform the best fit of the population dynamics to a single exponential decay and compute an ad-hoc rate constant. The total rate constants so computed are shown in Fig. \ref{fig_FSSH_rates} as a function of temperature and compared against analytical and AFSSH results. The analytical rate constants compare surprisingly well with the FSSH rate constants.  This is unexpected as the analytical rate constants are derived in the short time limit (linear regime) with decoherence included, while FSSH population dynamics is in the non-linear regime. Note that the maximum obtained in the rate constant at around 100 K is spurious -- exact HEOM results do not show this (Fig. 2 (c) of the main paper). We leave the correspondence of the FSSH rate constants with the analytical rate constants as an open problem for now.

\section{Simulation Details}

In this section, the simulation details of AFSSH, FSSH, and HEOM are discussed. 
For all dynamical calculations of the AFSSH method, we used 5000 trajectories with a classical time step of $0.25$ fs. For all FSSH calculations, we used 10000 trajectories with a classical time step of 0.2 fs, \mycomment{except for Fig. 3 of the main paper, where 1 million trajectories and a time step of 0.1 fs was used}. The initial position and momentum of each trajectory were obtained using the Boltzmann distribution. In the surface hopping algorithm, velocity reversal was done for a frustrated hop using Truhlar's velocity reversal scheme.\cite{jasper2003improved} The diabatic basis population calculation was done following the prescription given by Subotnik's group.\cite{subotnik2013can} The decoherence used is the augmented FSSH algorithm.\cite{landry2012recover}

To calculate the decoherence timescale $\tau$ for use in analytical formulae, we simulated 1000 trajectories for 10 ps for each parameter, and the average number of collapses was recorded during that time. The decoherence timescale was calculated as 10 ps / ``average number of collapses per trajectory." This timescale, shown in Table \ref{table_tau} for all parameters, was used in the analytical formula derived in this study. 

\begin{table}[H]
    \centering
    \begin{tabular}{c|c|c|c}
        $\Delta$G (cm$^{-1}$)& $\lambda$ (cm$^{-1}$)& T (K)& $\tau$ (fs)\\
        \hline
        -250 & 10 & 300 & 422.92 \\
        -275 & 10 & 300 & 450.76 \\
        -300 & 10 & 300 & 482.23 \\
        -325 & 10 & 300 & 517.66 \\
        -350 & 10 & 300 & 551.39 \\
        -375 & 10 & 300 & 588.69 \\
        -400 & 10 & 300 & 629.66 \\
        -350 & 5 & 300 & 824.57 \\
        -350 & 7.5 & 300 & 653.83 \\
        -350 & 10 & 300 & 551.39 \\
        -350 & 12.5 & 300 & 485.44 \\
        -350 & 15 & 300 & 433.20 \\
        -350 & 17.5 & 300 & 398.13 \\
        -350 & 20 & 300 & 369.71 \\
        -350 & 10 & 50 & 913.80 \\
        -350 & 10 & 100 & 755.93 \\
        -350 & 10 & 150 & 674.65 \\
        -350 & 10 & 200 & 620.00 \\
        -350 & 10 & 250 & 581.04 \\
        -350 & 10 & 300 & 551.26 \\
        -350 & 10 & 350 & 525.50 \\
        -350 & 10 & 400 & 504.05 \\
        -350 & 10 & 450 & 487.55 \\
        -350 & 10 & 500 & 471.75 \\
        -350 & 10 & 550 & 458.34 \\
        -350 & 10 & 600 & 444.92 \\
    \end{tabular}
    \caption{Values of the decoherence times used in the analytical formula for different parameters.}
    \label{table_tau}
\end{table}

To obtain numerically exact results, the HEOM calculations were converged by varying the time step, hierarchy depth (L), and number of Matsubara frequency (K) terms. For parameters with a 50 K temperature, the converged value of K is 2; for 100 K and 150 K, the converged value is 1, and for higher temperatures, it is 0. For all the parameters used, the converged value of L fell in the range of 6-12 and a time step of 0.05 fs. 

\bibliography{ref}